\documentstyle[12pt,pictex,psfig,epsf]{article}
\newcommand{\Tr}{\mbox{Tr}}
\newcommand{\Li}{\mbox{Li}}

\newcommand{\de}{\mbox{det}}
\newcommand{\del}{{\partial}}
\newcommand{\beq}{\begin{eqnarray}}
\newcommand{\eeq}{\end{eqnarray}}
\newcommand{\be}{\begin{eqnarray*}}
\newcommand{\ee}{\end{eqnarray*}}
\newcommand{\bk}{{\bf k}}

\newcommand{\br}{{\bf r}}
\newcommand{\bx}{{\bf x}}

\newcommand{\ra}{\rightarrow}
\newcommand{\e}{\epsilon}
\newcommand{\ve}{\varepsilon}
\newcommand{\nn}{\nonumber}

\newcommand{\wh}[1]{\widehat{#1}}
\newcommand{\ex}[1]{\langle\,#1\,\rangle}
\newcommand{\D}{{\cal D}}
\newcommand{\f}[2]{\mbox{$\frac{\scriptstyle#1}{\scriptstyle#2}$}}
\newcommand{\2}{\f{1}{2}}
\newcommand{\td}[1]{\bar{#1}}

\def\bg#1{\mbox{\boldmath$#1$}}

\begin{document}

\centerline{\Large\bf Thermodynamics of a Weakly Interacting}
\centerline{\Large\bf Bose-Einstein Gas}
\vskip 10mm
\centerline{T. Haugset, H. Haugerud and F. Ravndal}
\centerline{\it Department of Physics}
\centerline{\it University of Oslo}
\centerline{\it N-0316 Oslo, Norway}

\vspace{10mm}
{\bf Abstract:} {\footnotesize The one-loop effective potential for
non-relativistic bosons with a delta function repulsive potential is calculated 
for a given chemical potential using functional methods. After renormalization and 
at zero temperature it reproduces the standard ground state energy and pressure
as function of the particle density. At finite temperatures it is found
necessary to include ring corrections to the one-loop result in order to
satisfy the Goldstone theorem. It is natural to introduce an effective chemical
potential directly related to the order parameter and which uniformly decreases with
increasing temperatures.  This is in contrast to the the ordinary chemical potential
which peaks at the critical temperature. The resulting thermodynamics in the condensed
phase at very low temperatures is found to be the same as in the Bogoliubov approximation 
where the degrees of freedom are given by the Goldstone bosons. At higher temperatures 
the ring corrections dominate and result in a critical temperature unaffected by the
interaction.}

\section{Introduction}
Bose-Einstein condensation is central to much of our 
understanding of phenomena in condensed matter physics \cite{Huang}. It is 
one of the simplest processes where quantum effects manifest themselves on 
the macroscopic level when a finite fraction of the non-interacting bosons 
in a system start to occupy the lowest energy level. Although all the particles will 
then be in the same quantum state at zero temperature, this condensate does not have
real long-range order and does not truly represent a different phase.  It was
Bogoliubov \cite{Bogoliubov} who first showed that a short-range repulsion between the
particles is necessary in order to have a real condensate with the particles in a new
physical phase which is superfluid. His description of the condensation of interacting
bosons has since then formed the basis for a much more detailed understanding
of these important phenomena \cite{PN}.

Until very recently the only physical Bose-Einstein system of non-relativistic particles
exhibiting a phase transition at low temperatures, was liquid $He^4$. But here the particle
density is so high that it is a strongly interacting system so that perturbation
theory around the free system does not work \cite{PN}. However, with the recent
experimental progress made in connection with magnetically trapped bosons in the gas
phase \cite{BEC}, the situation has radically changed and systems of weakly interacting bosons
can now be studied. These were theoretically investigated in a series of papers by Lee, 
Yang and
their collaborators forty years ago using methods from statistical mechanics \cite{LeeYang_2,LeeYang_1,LHY,LeeYang_3}.
Their many results still represent to a large degree the state of theoretical understanding
of weakly interacting boson gases.

In the normal phase one can assign a definite number of particles to the 
system while this is impossible after the condensate has formed \cite{PWA}.
Bose-Einstein condensation of interacting particles is therefore the oldest and
still probably the simplest example of spontaneous symmetry breakdown which
today lies at the very heart of modern elementary particle theory \cite{Tony}.
Since these theories are relativistic, it is not obvious how Bogoliubov's
method can be used in this case. Instead one has developed very powerful
methods based on Feynman's path integral formulation of quantum field 
theories \cite{RPF} which allow the calculation of the corresponding
effective potentials in a very systematic way \cite{Bernard,DJ,Weinberg}.

This approach to spontaneous symmetry breakdown based upon functional
me\-thods has not yet been used to the same extent in the study of
Bose-Einstein condensation of non-relativistic systems although the basic 
elements are already in a modern textbook \cite{Brown}. In two dimensions 
it has been used by Lozano\cite{Lozano}. It is not clear
if this approach is equivalent to the Bogoliubov method or not. This is what
we have set out to investigate here. A first step in this direction was taken
several years ago by Kapusta \cite{Kapusta_1} who
considered interacting systems of relativistic bosons at non-zero chemical
potential and their condensation at low temperatures. More recently, Bernstein 
and Dodelson \cite{BD} and Benson, Bernstein  and Dodelson \cite{BBD} extended 
these relativistic calculations and also considered the non-relativistic 
limit. We will here show that their finite-temperature results are incomplete
in that they have not included the contributions from the ring or daisy
diagrams which are known to be essential at non-zero temperatures \cite{DJ,Weinberg,Carrington,AE}. Since then, the non-relativistic Bose gas has also 
been studied by Stoof and Bijlsma \cite{SB}.

The use of functional methods in quantum statistical physics of 
non-relativistic systems has not yet as widespread as for relativistic
systems. One of the best introductions have been given by Popov \cite{Popov}.
Our approach is different and more along the lines used in relativistic quantum
field theories \cite{Kapusta_2},  but we will to a large degree reproduce his results.
The necessary formalism is established in
the next section where we will derive the thermodynamics of a gas of free 
bosons in this language. We will work with the two real components of
the field instead of the complex field itself and its conjugate which is
usually done in condensed matter physics \cite{FW,Griffin}. We find this 
choice of variables especially advantageous in the case of interacting	
particles at zero temperature considered in Section 3. We calculate the
effective potential and free energy at non-zero chemical potential 
in the one-loop approximation where we
include the quantum effects of the fluctuations around the classical
solution. After removing the
divergences in the theory by renormalization of the coupling constant and
the chemical potential, we find the ground state energy of the hard-core 
bosons to be in exact agreement with the standard results of Lee and Yang
\cite{LeeYang_2}. In Section 4 we extend the calculation of the effective potential
to finite temperatures using the imaginary-time formalism. We discover that 
the one-loop approximation is no longer consistent with the Goldstone
theorem requiring the excitation spectrum to be linear in the long wavelength
limit. The problem is solved by including the so-called daisy or ring 
corrections to the boson propagator. These important contributions to the
free energy of relativistic quantum field theories at finite temperature 
were first discussed by Dolan and Jackiw \cite{DJ} and Weinberg \cite{Weinberg}. 
They have recently become
of importance in the connection with the standard model of elementary particles
at finite temperatures \cite{Carrington,AE}. In the last section we discuss the
obtained results and compare with what has been obtained by other methods.
This functional approach allows now in principle a systematic calculation of
higher loop corrections to the thermodynamics of the interacting boson gas.

\section{Functional methods for the non-relativistic boson gas}
The wavefunction $\psi = \psi(\bx,t)$ for a free, non-relativistic particle 
of mass $m$ satisfies the Schr\"odinger equation
\beq
     i\del_t\psi = - {1\over 2m}\nabla^2 \psi                 \label{free.1}
\eeq
when we use units so that $\hbar = 1$. In the second quantized description of 
a system of many such particles $\psi(\bx,t)$ becomes the corresponding quantum 
field. The wave equation (\ref{free.1}) is then the classical equation of 
motion. It follows from the Schr\"odinger Lagrangian
\beq
    {\cal L} = i\psi^*\del_t\psi 
             - {1\over 2m}|{\bg\nabla}\psi|^2                  \label{free.2}
\eeq
Constructing now the Hamiltonian $H$ and the number operator 
$N = \int\!d^3x\,\psi^*\psi$, the grand canonical partition function for the
is then given by
\beq
     \Xi(\beta,\mu) = \Tr\,e^{-\beta(H - \mu N)}               \label{free.3}
\eeq
where $\beta = 1/T$ when the system is in thermal equilibrium at temperature
$T$ and chemical potential $\mu$. The Boltzmann constant is taken to be
$k_B = 1$.	Rewriting now the trace as a path integral \cite{RPF}, one 
obtains the functional integral
\beq
   \Xi(\beta,\mu) = \int\!\D\psi\D\psi^*e^{-\int_0^\beta\!d\tau\!\int\!d^3x\,
                    {\cal L}_E(\psi^*,\psi)}                  \label{free.4}
\eeq
where the field $\psi = \psi(\bx,\tau)$ is now function of imaginary time 
$\tau = it$. Its dynamics is governed by the Euclidean Lagrangian density
\beq
    {\cal L}_E = \psi^*\del_{\tau}\psi + {1\over 2m}|{\bg\nabla}\psi|^2 
               - \mu\psi^*\psi                                \label{free.5}
\eeq
where we have included the contribution from the chemical potential.

\subsection{Complex field formalism}
Since the integral (\ref{free.4}) is Gaussian, it can be
immediately evaluated. The result for the free energy $\Omega = - \ln\Xi/\beta$
can then be written as
\beq
    \Omega(\beta,\mu) = {1\over\beta}\Tr\ln(\partial_\tau - {\nabla}^2/2m - \mu)
\eeq
The functional trace is given by eigenvalues of the indicated operator.
They are found by expanding the Bose field in plane waves
\beq
   \psi(\bx,\tau) = \sqrt{1\over\beta V} \sum_{n=-\infty}^\infty\sum_{\bk}
   \psi_{n\bk}\,e^{i\bk\cdot\bx + i\omega_n\!\tau}             \label{free.6}
\eeq
where $\omega_n = 2\pi n/\beta$ are the corresponding Matsubara frequencies.
Then we have 
\beq
    \Omega(\beta,\mu) = {1\over\beta}\sum_{\bk}\sum_{n=-\infty}^\infty
    \ln(-i\omega_n + \ve_\bk - \mu)                           \label{free.6b}
\eeq
where $\ve_\bk = \bk^2/2m$ is the single-particle energy.
One can regularize the divergent sum by taking the derivative with respect
to $e_\bk = \ve_\bk - \mu$. Using then the standard sum
\beq
    \sum_{n=-\infty}^\infty {1\over \omega_n^2 + \omega^2} 
           = {\beta\over\omega}\left[{1\over 2} 
           + {1\over e^{\beta\omega} - 1}\right]               \label{free.7}
\eeq
and integrating back, we have
\beq
    \Omega(\beta,\mu) = \sum_{\bk}\left[\2 e_\bk 
  + T\ln\left(1 - e^{-\beta e_\bk}\right)\right]               \label{free.8}
\eeq
after discarding an infinite constant. The pressure in the gas is now simply 
$P = -\Omega/V$. Taking the  infinite-volume limit, it becomes
\beq
    P(\beta,\mu) = - \int\!{d^3k\over(2\pi)^3}\left[\2 (\ve_\bk - \mu)
  + T\ln\left(1 - e^{-\beta(\ve_\bk - \mu)}\right)\right]      \label{free.9}
\eeq
The first term is the zero-point energy which is the only contribution at
zero temperature and is usually without physical content. However, in the 
next section where we discuss the interacting theory, it will be important. 

In order to find the equation of state, we want the pressure as a function
of the density	$\rho = \del P/\del\mu$. Ignoring the zero-point energy in 
(\ref{free.9}), we get the ordinary result
\beq
	  \rho = \int\!{d^3k\over(2\pi)^3}{1\over e^{\beta(\ve_\bk - \mu)} - 1} 
                                                                \label{be.1}
\eeq
from which one can calculate the chemical potential $\mu$ as function of 
temperature and density. From the pressure one then has the equation of state. It involves a critical line, determined by $\mu = 0$, i.e.
\beq
     \rho = \int\!{d^3k\over(2\pi)^3}{1\over e^{\beta\ve_\bk} - 1} 
          = \zeta(3/2) \left({mT\over 2\pi}\right)^{3/2}     \label{be.2}
\eeq
For densities above the critical value, the pressure in the gas is seen to
be independent of the density. This condensed phase has thus an infinite 
compressibility which shows that it is non-physical. 
A short-ranged, repulsive potential between the particles will solve this 
problem and the Bose-Einstein condensate will become a physical 
superfluid \cite{Huang}.

\subsection{Real field formalism}
When we later consider the interacting gas, we will find it more convenient
to take the two real components of the complex field $\psi$ as independent
field variables,
\beq
     \psi = \sqrt{1\over 2} (\psi_1 + i \psi_2)               \label{free.10}
\eeq
The Euclidean Lagrangian (\ref{free.5}) can then be written as
\beq
    {\cal L}_E = {1\over 2}\psi_a \wh{M}_{ab}\psi_b          \label{free.11}
\eeq
where equal indices are summed from 1 to 2. We have here introduced the 
matrix operator 
\beq
	 \wh{M}_{ab} = i\e_{ab}\del_\tau - \left({\nabla^2\over 2m} 
           + \mu\right)\delta_{ab}                            \label{free.12}
\eeq
where $\e_{ab}$ is the antisymmetric tensor in two dimensions with $\e_{12}=1$.
After expressing the corresponding action in terms of the complex Fourier components
in (\ref{free.6}), the partition function (\ref{free.4}) is now given by the
functional integral
\beq
   \Xi(\beta,\mu) = \int\!\D\psi_1\D\psi_2\,\exp{\left[-{1\over 2}\sum_{n,{\bk}}
   (e_{1\bk}\psi_1^*\psi_1  + e_{2\bk}\psi_2^*\psi_2) 
   + \sum_{n,{\bk}}\omega_n(\psi_1^*\psi_2 - \psi_2^*\psi_1)\right]}    \label{free.13}
\eeq
The two energies $e_{1\bk}$ and $e_{2\bk}$ are here actually both equal to
$e_\bk = \ve_\bk - \mu$. 
It should also be clear that in this integral we have 
suppressed the Fourier indices on the field components. 

From the form of the partition function (\ref{free.13}) we see that in 
terms of the real field components, the free, non-relativistic field theory 
is at the fundamental level interacting. The diagonal part of the action moves
the fields $\psi_1$ and $\psi_2$ at fixed time with the propagators
\beq
     \setcoordinatesystem units <1cm,1cm> 
     \unitlength=1cm
     \plot -0.9 0.12 -0.1 0.12 /
     : D_{11}^{(0)} = {1\over e_{1\bk}} \hspace{3.5cm}
     \setdashes <2pt>
     \plot -0.9 0.12 -0.1 0.12 /
     : D_{22}^{(0)} = {1\over e_{2\bk}}                   \label{free.14}
\eeq
while the kinetic term provides the interaction with the simple vertex
\beq
    \setcoordinatesystem units <1cm,1cm> 
    \unitlength=1cm
    \plot -1.0 0.12 -0.2 0.12 /
    \put {\circle{0.2}} [B1] at 0.025 0.12
    \setdashes <2pt>
    \plot 0.05 0.12 0.85 0.12 /
    \hspace{1cm}
    : \omega_n                   \hspace{3.5cm}                
    \plot -1.0 0.12 -0.2 0.12 /
    \put {\circle{0.2}} [B1] at 0.025 0.12
    \setsolid 
    \plot 0.05 0.12 0.85 0.12 /
    \hspace{1cm}
    : -\omega_n         \label{free.15}
\eeq
of strength given by the Matsubara frequency. The full, free energy is now
obtained in standard perturbation theory which is almost trivially solved
to all orders in the interaction. First, we need the partition function of
the free theory
\beq
   \Xi_0 & = & \int\!\D\psi_1\D\psi_2\,\exp{\left[-{1\over 2}\sum_{n,{\bk}}
   (e_{1\bk}\psi_1^*\psi_1  + e_{2\bk}\psi_2^*\psi_2)\right]} \nn \\
   & = & \de^{-\2}(e_{1\bk})\,\de^{-\2}(e_{2\bk})		\label{free.16}
\eeq
Taking the logarithm, we find the non-interacting result
\beq
   \beta\Omega_0 & = & {1\over 2}\left[\Tr\ln e_{1\bk} + \Tr\ln e_{2\bk}\right] \nn\\
                 & = &  -\left[\frac{1}{2}\ln
	\setcoordinatesystem units <1cm,1cm> 
	\unitlength=1cm
	\circulararc 360 degrees from 0.9 0.12 center at 0.5 0.12
	\hspace{1cm}
	+ \frac{1}{2}\ln
	\setdashes <1.50pt>
	\circulararc 360 degrees from 0.9 0.12 center at 0.5 0.12 
	\hspace{1.00 cm} \right]
                                                            \label{free.17}
\eeq
where the closed loops denotes the trace over the variables in the propagators.
The kinetic interaction will now perturb the fields in these two 
loop diagrams. A 1-field will be converted two a 2-field and vice versa with the coupling constant $\omega_n$. Since the free propagator 
$\ex{\psi_1\psi_2}_0 = 0$, only loops with an even number of interactions will 
contribute, i.e. with the same number of 1- and 2-fields. We then find for
the full free energy
\beq
\beta\Omega & = & -\left[\frac{1}{2}\ln
\setcoordinatesystem units <1cm,1cm> 
\unitlength=1cm
\circulararc 360 degrees from 0.9 0.12 center at 0.5 0.12
\hspace{1cm}
+ \frac{1}{2}\ln
\setdashes <1.50pt>
\circulararc 360 degrees from 0.9 0.12 center at 0.5 0.12
\hspace{1.00 cm}
+ \frac{1}{2}
\put {\circle{0.16}} [Bl] at 0.50 0.52
 \setsolid
\circulararc 156 degrees from 0.412 0.510 center at 0.500 0.120
\put {\circle{0.16}} [Bl] at 0.50 -0.28
 \setdashes <1.50pt>
\circulararc 156 degrees from 0.588 -0.270 center at 0.500 0.120
\hspace{1.00 cm}
+ \frac{1}{4}
\put {\circle{0.16}} [Bl] at 0.50 0.52
 \setsolid
\circulararc 66 degrees from 0.412 0.510 center at 0.500 0.120
\put {\circle{0.16}} [Bl] at 0.10 0.12
 \setdashes <1.50pt>
\circulararc 66 degrees from 0.110 0.032 center at 0.500 0.120
\put {\circle{0.16}} [Bl] at 0.50 -0.28
 \setsolid
\circulararc 66 degrees from 0.588 -0.270 center at 0.500 0.120
\put {\circle{0.16}} [Bl] at 0.90 0.12
 \setdashes <1.50pt>
\circulararc 66 degrees from 0.890 0.208 center at 0.500 0.120
\hspace{1.00 cm}
+ \frac{1}{6}
\put {\circle{0.16}} [Bl] at 0.50 0.52
 \setsolid
\circulararc 36 degrees from 0.412 0.510 center at 0.500 0.120
\put {\circle{0.16}} [Bl] at 0.15 0.32
 \setdashes <1.50pt>
\circulararc 36 degrees from 0.118 0.239 center at 0.500 0.120
\put {\circle{0.16}} [Bl] at 0.15 -0.08
 \setsolid
\circulararc 36 degrees from 0.206 -0.151 center at 0.500 0.120
\put {\circle{0.16}} [Bl] at 0.50 -0.28
 \setdashes <1.50pt>
\circulararc 36 degrees from 0.588 -0.270 center at 0.500 0.120
\put {\circle{0.16}} [Bl] at 0.85 -0.08
 \setsolid
\circulararc 36 degrees from 0.882 0.001 center at 0.500 0.120
\put {\circle{0.16}} [Bl] at 0.85 0.32
 \setdashes <1.50pt>
\circulararc 36 degrees from 0.794 0.391 center at 0.500 0.120
\hspace{1.00 cm}
+ \cdots\right]
\nn\\
   & = & {1\over 2}\sum_{n,\bk}\left[\ln(e_{1\bk}e_{2\bk})
     + {\omega_n^2\over 1(e_{1\bk}e_{2\bk})} 
     - {\omega_n^4\over 2(e_{1\bk}e_{2\bk})^2}
     + {\omega_n^6\over 3(e_{1\bk}e_{2\bk})^3} + \cdots\right] \nn\\
   & = & {1\over 2}\sum_{n,\bk}\ln(\omega_n^2 + e_{1\bk}e_{2\bk}) 
                                                            \label{free.18}
\eeq
Again regularizing as in ({\ref{free.6b}) and summing over the Matsubara 
frequencies with the help of (\ref{free.7}), we recover the the standard
Bose-Einstein free energy (\ref{free.8}).

\subsection{Free propagators of real fields}
The propagators (\ref{free.14}) move only the fields at fixed time. Motion
in time is induced by the kinetic interaction in (\ref{free.11}). Its full effect
can easily be calculated in perturbation theory. For the 1-field, when
we again consider a Fourier mode with a given momentum $\bk$ and Matsubara
energy $\omega_n$, we find:

\beq 
\setcoordinatesystem units <1cm,1cm> 
\unitlength=1cm
   D_{11} & = & \ex{\psi_1\psi_1^*} = 
  		\plot 10 2.8 50 2.8 / 		
  		\plot 10 3.0 50 3.0 / 		
		\plot 10 3.2 50 3.2 /  
		\plot 10 3.4 50 3.4 /  
		\plot 10 3.6 50 3.6 / 
  		\plot 10 3.8 50 3.8 /\nonumber 	\\
          & = & \plot 5 3.3 25 3.3 /  \hspace{1.05cm} +
		\plot 5 3.3 25 3.3 /
		\put {\circle{6}} [B1] at 32 3.3
		\setdashes <2pt>
		\plot 32 3.3 53 3.3 / 
		\put {\circle{6}} [B1] at 59 3.3
		\setsolid
		\plot 59 3.3 79 3.3 / \hspace{2.95cm} + 
		\plot 5 3.3 25 3.3 /
		\put {\circle{6}} [B1] at 32 3.3
		\setdashes <2pt>
		\plot 32 3.3 53 3.3 / 
		\put {\circle{6}} [B1] at 59 3.3
		\setsolid
		\plot 59 3.3 79 3.3 /
		\put {\circle{6}} [B1] at 86 3.3
		\setdashes <2pt>
		\plot 86 3.3 107 3.3 / 
		\put {\circle{6}} [B1] at 113 3.3
		\setsolid
		\plot 113 3.3 133 3.3 /
		\hspace{4.8cm} + \cdots\nonumber\\
	   & = & D_{11}^{(0)} + D_{11}^{(0)}(\omega_n)D_{22}^{(0)}(-\omega_n)D_{11}^{(0)}\nonumber \\
           & + & D_{11}^{(0)}(\omega_n)D_{22}^{(0)}(-\omega_n)D_{11}^{(0)}(\omega_n)D_{22}^{(0)}(-\omega_n)D_{11}^{(0)} + \cdots\nonumber \\
 	   & = & {D_{11}^{(0)}\over 1 + \omega_n^2D_{11}^{(0)}D_{22}^{(0)}}
             = {e_{2\bk}\over\omega_n^2 + e_{1\bk}e_{2\bk}} 
\eeq
Similarly, we find
\beq 
    \setcoordinatesystem units <1cm,1cm> 
    \unitlength=1cm
    D_{22} & = & \ex{\psi_2\psi_2^*} = 
		\setdashes <2pt>
  		\plot 10 2.6 50 2.6 / 		
  		\plot 10 2.8 50 2.8 / 		
  		\plot 10 3.0 50 3.0 / 		
		\plot 10 3.2 50 3.2 /  
		\plot 10 3.4 50 3.4 /  
		\plot 10 3.6 50 3.6 / 
 		\plot 10 3.8 50 3.8 /\nonumber \\
          & = & \setdashes <2pt>
		\plot 5 3.3 25 3.3 /  \hspace{1.05cm} +
		\plot 5 3.3 26 3.3 /
		\put {\circle{6}} [B1] at 32 3.3
		\setsolid
		\plot 32 3.3 53 3.3 / 
		\put {\circle{6}} [B1] at 59 3.3
		\setdashes <2pt>
		\plot 59 3.3 79 3.3 / \hspace{2.95cm} + 
		\plot 5 3.3 26 3.3 /
		\put {\circle{6}} [B1] at 32 3.3
		\setsolid
		\plot 32 3.3 53 3.3 / 
		\put {\circle{6}} [B1] at 59 3.3
		\setdashes <2pt>
		\plot 59 3.3 80 3.3 /
		\put {\circle{6}} [B1] at 86 3.3
		\setsolid
		\plot 86 3.3 107 3.3 / 
		\put {\circle{6}} [B1] at 113 3.3
		\setdashes <2pt>
		\plot 113 3.3 133 3.3 /
		\hspace{4.8cm} + \cdots\nonumber\\
	   & = & {e_{1\bk}\over\omega_n^2 + e_{1\bk}e_{2\bk}} 
\eeq
and
\beq
\setcoordinatesystem units <1cm,1cm> 
\unitlength=1cm
	 D_{12} & = & \ex{\psi_1\psi_2^*} = 
  		\plot 10 2.8 30 2.8 / 		
  		\plot 10 3.0 30 3.0 / 		
		\plot 10 3.2 30 3.2 /  
		\plot 10 3.4 30 3.4 /  
		\plot 10 3.6 30 3.6 /  
		\plot 10 3.8 30 3.8 / 
		\thicklines
		\put {\circle{6}} [B1] at 36 3.3 
		\setdashes <2pt>
  		\plot 36 2.8 56 2.8 / 		
  		\plot 36 3.0 56 3.0 / 		
		\plot 36 3.2 56 3.2 /  
		\plot 36 3.4 56 3.4 /  
		\plot 36 3.6 56 3.6 /  
		\plot 36 3.8 56 3.8 /\nonumber \\
          & = & \setsolid
		\plot 5 3.3 25 3.3 /
		\put {\circle{6}} [B1] at 32 3.3
		\setdashes <2pt>
		\plot 32 3.3 52 3.3 / \hspace{2cm} +  
		\setsolid
		\plot 5 3.3 25 3.3 /
		\put {\circle{6}} [B1] at 32 3.3
		\setdashes <2pt>
		\plot 32 3.3 53 3.3 / 
		\put {\circle{6}} [B1] at 59 3.3
		\setsolid
		\plot 59 3.3 79 3.3 /
		\put {\circle{6}} [B1] at 86 3.3
		\setdashes <2pt>
		\plot 86 3.3 106 3.3 / 
		\hspace{4cm} + \cdots\nonumber \\
           & = & {\omega_n\over\omega_n^2 + e_{1\bk}e_{2\bk}} 
\eeq
while $D_{21} = -D_{12}$. These results can simply be summed up in the
Dyson-Schwinger equations for the full propagators
\beq
	 D_{ab} = D_{ab}^{(0)} + D_{ac}^{(0)}\Pi_{cd}^{(0)}D_{db}   \label{prop.5}
\eeq
With the free propagator (\ref{free.14}), $D_{12}^{(0)} = 0$ and the 
non-diagonal self energy $\Pi_{cd}^{(0)} = \e_{cd}\omega_n$, the equations
are easily solved to give the same results as above. 

We will later use the Schwinger-Dyson equations to calculate the propagators 
when the bosons have a short-range interaction. Here
the fields are essentially free and the propagators can be obtained directly
from the matrix operator (\ref{free.12}). Its Fourier transform is
\beq
     M_{ab} = \left(\begin{array}{rr} e_{1\bk}  &  -\omega_n  \\
               	                 \omega_n  &    e_{2\bk} \end{array}\right)
\eeq
Taking the inverse, we then simply have
\beq
    D_{ab} =  \ex{\psi_a\psi_b} = M_{ab}^{-1}          
           =  {1\over\omega_n^2 + e_{1\bk}e_{2\bk}}\left(
           \begin{array}{rr} e_{2\bk}  &  \omega_n  \\
          -\omega_n  &    e_{1\bk} \end{array}\right)        \label{prop.6}
\eeq
which is seen to agree with the previous results.

\section{Hard-core bosons in the one-loop approximation}
We will here consider the idealized case of bosons having only a repulsive
interaction potential $V(\br)$ at short distances. The thermodynamics of the 
gas will then be mostly independent of the detailed shape of the potential
which will only enter the results via the $S$-wave scattering length \cite{PN}
\beq
     a = {m\over 4\pi}\int\!d^3r\, V(\br)                       \label{int.1}
\eeq
which is positive. This is equivalent to saying that the potential is a 
$\delta$-function, i.e. $V(\br) = 2\lambda\,\delta(\br)$ with the coupling 
constant $\lambda = 2\pi a/m$.
In the second quantized theory it will correspond to an interaction term
$\lambda(\psi^*\psi)^2$ in the Lagrangian. While the coupling constant 
$\lambda$ would be dimensionless in the corresponding relativistic theory,
it is not in the non-relativistic description we use here. Let us comment
briefly upon this point.

The Lagrangian of a real relativistic scalar field $\Psi$
can be written as
\beq
{\cal L}_0 = \frac{1}{2}\del_\mu\Psi\del^\mu\Psi - \frac{m^2}{2}\Psi^2
- \lambda_0\Psi^4
\eeq
in real time formalism.
Here $\mu = 0\ldots 3$ and $\lambda_0$ is the 
relativistic coupling constant. Since the action must be dimensionless,
the field takes on the same dimension as the inverse time or inverse
distance in units where the velocity of light $c=1$.
The coupling $\lambda_0$ is therefore dimensionless. 
We now take the non-relativistic limit by letting $m\rightarrow\infty$. Before
doing so, we introduce the non-relativistic field $\psi$ through
\beq
\Psi = \frac{1}{\sqrt{2m}}\left(e^{-imt}\psi + e^{imt}\psi^*\right)
\eeq
This leads to a number of terms in the Lagrangian which oscillate with 
frequency $2m$. They may be dropped as $m\rightarrow\infty$. The resulting
non-relativistic Lagrangian takes the form
\beq
{\cal L} = i\psi^*\del_t\psi - \frac{1}{2m}|{\bg\nabla}\psi|^2 
               - \lambda(\psi^*\psi)^2
\eeq
with a non-relativistic coupling constant $\lambda = 3\lambda_0/2m^2$. 
This coupling is obviously not dimensionless.

\subsection{The classical ground state}
Including the above interaction, the {\em Euclidean} Lagrangian (\ref{free.5}) 
describing the bosons is changed into
\beq
    {\cal L}_E = \psi^*\del_{\tau}\psi + {1\over 2m}|{\bg\nabla}\psi|^2 
               - \mu\psi^*\psi + \lambda(\psi^*\psi)^2         \label{int.3}
\eeq
In the classical limit at zero temperature the system will be in the lowest
energy state. The field will then attain a constant value given by the
minimum of the classical potential
\beq
    U(\psi) = - \mu\psi^*\psi + \lambda(\psi^*\psi)^2			\label{int.4}
\eeq
It is invariant under the $U(1)$ phase transformation $\psi(\bx) 
\ra e^{i\theta} \psi(\bx)$
and thus depends only on the modulus $|\psi|$ of the field. In Fig.1 the classical potential is
plotted for the two cases $\mu > 0$ and $\mu < 0$. We see that in the first case, we will have a ground state with spontaneous breakdown
of the $U(1)$ symmetry in which the field takes the classical value
$|\psi| = \sqrt{\mu/2\lambda}$. We will in later sections see that $\mu$ will be negative at high temperatures and the gas 
will be in the normal state with $|\psi| = 0$.
\begin{figure}[htb]
\begin{center}
\mbox{\psfig{figure=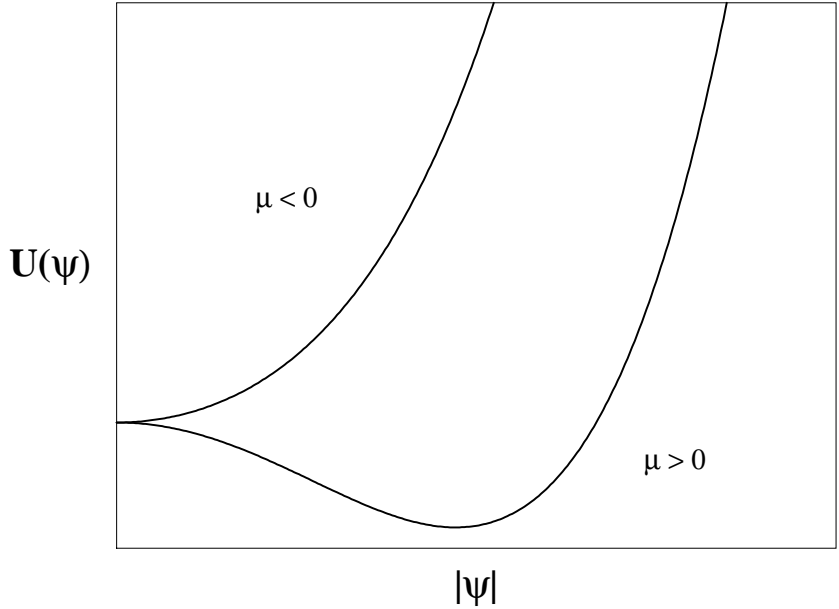,width=11cm,angle=270,height=8cm}}
\end{center}
\caption[The classical potential $U(\psi)$ plotted as function of the field modulus $|\psi|$ in arbitrary units.]{\footnotesize The classical potential $U(\psi)$ plotted as function of the field modulus $|\psi|$ in arbitrary units.}
\label{bec3D-fig:class}
\end{figure}

To be more quantitative, the symmetry-broken ground state has a classical
pressure given by the minimum of the potential (\ref{int.4}), i.e. 
$P(\mu) = \mu^2/4\lambda$. The corresponding number density is then $\rho = \mu/2\lambda$
and therefore $P = \lambda\rho^2$. This is also the classical ground state 
energy density ${\cal E}(\rho)$ as follows from the thermodynamic Legendre 
transform
\beq
		 {\cal E}(\rho) = \mu\rho - P(\mu)                      \label{int.5}
\eeq
In this way we have already cured one of the problems of the ideal Bose-Einstein
gas, namely the infinite compressibility of the condensed phase. What is needed
next is the inclusion of quantum and thermal fluctuations around this
classical ground state.

\subsection{Gaussian fluctuations}

Denoting the fluctuating field by $\chi$, we can write the full Bose field as
\beq
     \psi(x) = \sqrt{\2}\,[v + \chi(x)]                 \label{fluct.1}
\eeq
where $v$ is the constant field of the condensate. Inserting this into the Lagrangian (\ref{int.3}),
we can rewrite it as 	
\beq
    {\cal L}_E = - {\mu\over 2}|v|^2 + {\lambda\over 4}|v|^4
                 + {1\over 2}\chi_a\wh{M}_{ab}\chi_b 
                 + \lambda (v\cdot\chi)(\chi\cdot\chi)
                 + {\lambda\over 4}(\chi\cdot\chi)^2		  \label{fluct.2}
\eeq
when expressed in terms of the two real components of the complex fields 
$\chi$ and $v$. Terms linear in $\chi$ have been dropped since they will not contribute
after integration over space. The matrix operator $\wh{M}_{ab}$ is now
\beq
	 \wh{M}_{ab} = i\e_{ab}\del_\tau - \left({\nabla^2\over 2m} -\lambda(v\cdot v)
                     + \mu\right)\!\delta_{ab} + 2\lambda\,v_av_b               \label{fluct.3}
\eeq
Due to the $U(1)$ symmetry of the system, we can choose the classical field
$ v$ to be real. Taking the Fourier transform of the operator, it becomes 
\beq
     M_{ab} = \left(\begin{array}{cc} \ve_\bk - \mu + 3\lambda v^2  &  -\omega_n  \\
              \omega_n  & \ve_\bk - \mu + \lambda v^2  \end{array}\right)
                                                             \label{fluct.4}
\eeq
where again $\ve_\bk = \bk^2/2m$.

The grand canonical partition function (\ref{free.4}) can be now written as a
functional integral over the real fields $\chi_1$ and $\chi_2$ as
\beq
     \Xi(\beta,\mu) = e^{-\beta V(-{\mu\over 2}v^2 + {\lambda\over 4}v^4)}
     \int\!\D\chi_1\D\chi_2\,e^{-\int_0^\beta\!d\tau\!\int\!d^3x\,
     [{1\over 2}\chi_a\wh{M}_{ab}\chi_b + {\cal L}_{int}]}	         \label{fluct.5}
\eeq
where $V$ is the volume of the system and
\beq
     {\cal L}_{int} = \lambda v\chi_1(\chi\cdot\chi)
     + {\lambda\over 4}(\chi\cdot\chi)^2                               \label{fluct.5b}
\eeq
gives the interactions of the fluctuating field. The contribution from the first
part of the Lagrangian being quadratic in the fields, can be evaluated as for the
free theory in Section 2 and will be the one-loop result. Higher loop corrections 
due to the cubic and quartic terms in ${\cal L}_{int}$ can then be systematically 
calculated in perturbation theory. These will be especially important at finite temperatures
and will be considered in the next section.

In the one-loop approximation, we keep only the quadratic part of the Lagrangian.
The integral (\ref{fluct.5}) is then Gaussian and gives the corresponding
free energy
\beq
    {1\over V}\,\Omega(\mu,v) = -{\mu\over 2}v^2 + {\lambda\over 4}v^4
    + {1\over 2\beta V}\ln\de M					 \label{fluct.6} 
\eeq
From (\ref{fluct.4}) we find the determinant $\de M = \prod_{n,\bk}(\omega_n^2 
+ \omega_\bk^2)$ where the excitation energy is now
\beq
     \omega_\bk = \sqrt{(\ve_\bk - \mu + \lambda v^2)
                    (\ve_\bk - \mu + 3\lambda v^2)}              \label{fluct.7}
\eeq                                                    
Summing over the Matsubara frequencies using (\ref{free.7}) again, we obtain 
for the free energy (\ref{fluct.6})
\beq
    {1\over V}\,\Omega(\mu,v) = -{\mu\over 2}v^2 + {\lambda\over 4}v^4
    + {1\over V}\sum_\bk \left[{1\over 2}\omega_\bk 
    + T\ln\left(1 - e^{-\beta\omega_\bk}\right)\right]       \label{fluct.8}
\eeq
As a function of the condensate $v$, this is the effective potential $U_{eff}$ for the 
system in the one-loop approximation. The thermodynamical free energy is the value of the 
function in its minimum and equals the negative of the pressure $P(\mu,T)$. Due
to the quantum fluctuations, this will be slightly shifted away from the classical
minimum at $v^2 = \mu/\lambda$. For this value of the condensate, the dispersion
relation (\ref{fluct.7}) simplifies to the Bogoliubov result \cite{Bogoliubov}
\beq
     \omega = \sqrt{\ve(\ve + 2\mu)}
            = {k\over 2m}\sqrt{k^2 + 4m\mu}                  \label{fluct.9}
\eeq
Here, and in the following we drop the wave number index on $\omega$. In the long-wavelength limit $k\ra 0$ it becomes linear and represents the phonon excitations. This is
a consequence of the Goldstone theorem which requires excitations with a 
linear dispersion relation whenever a continuous symmetry is spontaneously 
broken as here \cite{PWA}. We will find in Section 4 that the
quantum corrections will not change this dispersion relation except that the
classical chemical potential $\mu$ is replaced by an effective potential $\td{\mu}$
which results from the quantum self energies in the field propagators.

\subsection{Renormalization}

The infinite sum (\ref{fluct.8}) is seen to be strongly divergent in the ultraviolet 
limit $k \ra \infty$. A similar divergence was also found in the free theory in 
Section 2. We can thus remove the strongest divergence by subtracting the first term 
in (\ref{free.8}) so that we recover the pressure in the non-interacting gas. Taking the
infinite-volume limit we then have for the effective potential
\beq
    U_{eff}(v,\mu,T) = -{\mu\over 2}v^2 + {\lambda\over 4}v^4
    + \int\!{d^3k\over(2\pi)^3}\!\left[{1\over 2}(\omega - \ve + \mu) 
    + T\ln\left(1 - e^{-\beta\omega}\right)\right]       \label{fluct.8b}
\eeq
However, the integral representing the Gaussian fluctuations at zero temperature is 
still not finite. The divergences can be removed
by renormalizing the coupling constant $\lambda$ and the chemical potential $\mu$.
For this purpose we introduce the counter-terms
\beq
    {\cal L}_{ct} & = &- \delta\mu\psi^*\psi + \delta\lambda(\psi^*\psi)^2 
                     = - {1\over 2}\delta\mu\,v^2 + {1\over 4}\delta\lambda\,v^4 
             + {1\over 4}\delta\lambda(\chi\cdot\chi)^2  \nn \\
           & - & {1\over 2}(\delta\mu - 3v^2\delta\lambda)\chi_1^2 
             - {1\over 2}(\delta\mu -  v^2\delta\lambda)\chi_2^2               \label{ct.1}
\eeq
resulting directly from the classical potential (\ref{int.4}). Again we have ignored terms
linear in the fluctuating field $\chi$. At zero temperature we then get for the renormalized 
effective potential
\beq
     U_{eff}(v,\mu,T=0) & = & -{\mu\over 2}v^2 + {\lambda\over 4}v^4 
      \nonumber\\
     & + & {1\over 2}\int\!{d^3k\over(2\pi)^3}\!\left(\omega - \ve + \mu\right)
       - {1\over 2}\delta\mu\,v^2 
       + {1\over 4}\delta\lambda\,v^4                             \label{re.2}
\eeq
This can now be made finite by adjusting the quantities $\delta\mu$ and $\delta\lambda$.
The temperature-dependent contribution in (\ref{fluct.8b}) is finite by itself and thus
the full free energy will be finite.

Explicit expressions for the counter-terms can be obtained by isolating the two 
leading divergences in the integral (\ref{re.2}) by expanding the expression 
(\ref{fluct.7}) for $\omega$ for large values of the momentum $k$. Cutting off
the integral at $k = \Lambda$, we can then rewrite $U_{eff}$ as
\be
    && U_{eff}(v,\mu,T=0)  = -{\mu\over 2}v^2 + {\lambda\over 4}v^4 
      - {1\over 2}\delta\mu\,v^2  + {1\over 4}\delta\lambda\,v^4 \\
     & + & {1\over 2}\int^\Lambda\!{d^3k\over(2\pi)^3}\!
     \left(\omega - \ve + \mu - 2\lambda v^2 + {\lambda^2v^4\over 2\ve}\right)
     + {1\over 2}\int^\Lambda\!{d^3k\over(2\pi)^3}\!
     \left(2\lambda v^2 - {\lambda^2v^4\over 2\ve}\right)                \label{re.3}
\ee
The divergences in the last integral are removed by taking the counter-terms to be
\beq
     \delta\mu = \int^\Lambda\!{d^3k\over(2\pi)^3}\,2\lambda 
               = {\lambda\over 3\pi^2}\Lambda^3                          \label{re.4}
\eeq
and
\beq
     \delta\lambda = \int^\Lambda\!{d^3k\over(2\pi)^3}{\lambda^2\over\ve} 
               = {\lambda^2\over\pi^2}m\Lambda                            \label{re.5}
\eeq
These give the same renormalized parameters as obtained by Benson, Bernstein
and Dodelson \cite{BBD}. Jackiw \cite{Jackiw} has also obtained the same
renormalized coupling constant from considerations of bound states of 
non-relativistic particles in a $\delta$-function potential.
The unrenormalized coupling constant $\lambda_0$ is thus given in terms of the 
cut-off by
\beq
    {1\over\lambda_0} = {1\over\lambda} - m{\Lambda\over\pi^2}            \label{re.6}
\eeq
For fixed, renormalized coupling $\lambda$ it increases with the cut-off. In the
corresponding relativistic theory, this increases is logarithmic instead of linear.

Having removed the divergences, we can now let the cut-off $\Lambda\ra\infty$. 
We then have the finite result
\beq
     U_{eff}(v,\mu,T=0) & = & -{\mu\over 2}v^2 + {\lambda\over 4}v^4 \nonumber\\
     & + & {1\over 2}\int\!{d^3k\over(2\pi)^3}\!
     \left(\omega - \ve + \mu - 2\lambda v^2 + {\lambda^2v^4\over 2\ve}\right)
                                                                           \label{re.7}
\eeq
With definite values for the couplings $\mu$ and $\lambda$ it can be calculated by a
numerical integration. As a function of the condensate value $v$ it has the same
shape as in Fig. 1. It will have an 
imaginary part in the region below the classical minimum at $v^2 = \mu/\lambda$ 
whose physical significance has been discussed by Weinberg and Wu \cite{WW}.

\subsection{Condensate and pressure at zero temperature} 

Because of the quantum fluctuations, the minimum of the effective potential (\ref{re.7})
is shifted away from the classical value. In the new minimum the derivative\\ 
$(\del U_{eff}/\del v)_\mu = 0$. With
\beq
    \left({\del\omega\over\del v}\right)_\mu = (2\ve - 2 \mu +3\lambda v^2){1\over\omega}
\eeq
we find the minimum to be at the condensate value
\beq
    v_0^2 = {\mu\over\lambda} - \int\!{d^3k\over(2\pi)^3}\!
          \left({2\ve - 2 \mu +3\lambda v^2\over\omega} - 2 + {\lambda v^2\over\ve}\right)
                                                                         \label{re.9}
\eeq
The integral represents here the effects of the quantum fluctuations. To this order in
perturbation theory we can evaluate it using the classical value $v^2 = \mu/\lambda$
for the condensate. It then gives
\beq
    v_0^2 = {\mu\over\lambda} - \int\!{d^3k\over(2\pi)^3}\!
          \left({2\ve +\mu\over\omega} +{\mu\over\ve} - 2\right)            \label{re.10}
\eeq
where now the Bogoliubov frequency $\omega$ is given by (\ref{fluct.9}). We will in the
next section show that the quantum shift of the classical minimum is due to the 
self energies of the interacting fields. These will effectively change the chemical 
potential from the classical value $\mu$ to a new value $\td{\mu}$ which is just 
$\lambda$ times the right-hand side of equation (\ref{re.10}) at zero temperature.

With the above value for the condensate in the minimum of the effective potential, we 
find by insertion into (\ref{re.7}) the value for the thermodynamic pressure at zero 
temperature. To lowest order in the quantum correction, it can be written as
\beq
    P(\mu) & = & {\mu^2\over 4\lambda} - {1\over 2}\int\!{d^3k\over(2\pi)^3}\!
    \left(\omega - \ve - \mu + {\mu^2\over 2\ve}\right)            \label{re.11}
\eeq
The full density of particles is now given by the derivative $\rho=\del P/\del\mu$,
i.e.
\beq
    \rho = {\mu\over 2\lambda} - {1\over 2}\int\!{d^3k\over(2\pi)^3}\!
          \left({\ve\over\omega} + {\mu\over\ve} - 1\right)         \label{re.12}
\eeq
Most of the particles are in the condensate with $\bk = 0$. Their density $\rho_c = 
v_0^2/2$ follows directly from (\ref{re.10}). The density $\rho_e = \rho - \rho_c$ of
particles in excited states with $\bk \ne 0$ is thus
\beq
    \rho_e = {1\over 2}\int\!{d^3k\over(2\pi)^3}\!
          \left({\ve + \mu\over \omega}  - 1\right)                    \label{re.13}
\eeq
It is caused by the hard-core repulsion between the particles and was also obtained by
Benson, Bernstein and Dodelson \cite{BBD}.

With the Bogoliubov dispersion relation (\ref{fluct.9}) for the excitation energy
$\omega$, we can now evaluate the pressure in (\ref{re.11}). Besides elementary 
integrations, it involves the integral
\beq
    \int\!d\ve \,\ve\sqrt{\ve + 2\mu} = {2\over 5}(\ve + 2\mu)^{5/2}
    - {4\over 3}\mu (\ve + 2\mu)^{3/2}                                \label{re.14}
\eeq
By construction, we now get only a non-zero contribution from the lower limit 
$\ve = 0$ of the integrals which gives
\beq
    P(\mu)  = {\mu^2\over 4\lambda} - {8m^{3/2}\over 15\pi^2}\mu^{5/2}
                                                                   \label{re.15}
\eeq
The last term represents the one-loop, quantum corrections to the classical
result in the first term. This agrees with the original results of Lee and
Yang \cite{LeeYang_1} who considered the same system of interacting bosons with
a hard-core repulsion within the framework of quantum statistical mechanics.

Equation (\ref{re.12}) enables us to relate the chemical potential to the 
particle density. It can be obtained more directly by just taking the derivative
of (\ref{re.15}) with respect to $\mu$,
\beq
    \rho  = {\mu\over 2\lambda} - {4\over 3\pi^2}(m\mu)^{3/2}    \label{re.16}
\eeq
The sign of the second term here is apparently opposite to what was obtained by
Bernstein and Dodelson \cite{BD}. It must be negative as here in order to have an increase in the zero-temperature pressure because of the repulsion between the particles and not. By inversion, we then obtain for the chemical potential to lowest order in perturbation theory,
\beq
    \mu = 2\lambda\rho + {8\lambda\over 3\pi^2}(2m\lambda\rho)^{3/2}
                                                                 \label{re.17}
\eeq
Similarly, from (\ref{re.13}) we find the density $\rho_e = (m\mu)^{3/2}/3\pi^2$ 
of particles in excited states. Expressed instead in terms of the full density,
it is
\beq
    \rho_e = {1\over 3\pi^2}(2m\lambda\rho)^{3/2} 
           =  {8\rho\over 3}\sqrt{\rho a^3\over\pi}             \label{re.18}
\eeq
when the coupling constant $\lambda$ is replaced by the scattering length $a$. This is just the textbook result \cite{Huang}.

With the chemical potential from (\ref{re.17}), we now get for the pressure (\ref{re.15})
\beq
    P(\rho) & = &\lambda\rho^2 + {4m^{3/2}\over 5\pi^2}(2\lambda\rho)^{5/2}
                                                                             \label{re.19}
\eeq 
The corresponding energy density follows then from the Legendre transformation 
${\cal E} = \mu\rho - P$ and is
\beq
    {\cal E}(\rho) = \lambda\rho^2 
                   + {8m^{3/2}\over 15\pi^2}(2\lambda\rho)^{5/2}         \label{re.19b}
\eeq
It was first calculated by Lee and Yang \cite{LeeYang_2} using the binary
collision method and also by Lee, Huang and Yang \cite{LHY} who used instead
the pseudopotential method. Usually, it is expressed in terms of the
scattering length and takes then the form
\beq
    {\cal E} = {2\pi a\over m}\rho^2 \left[1 
             + {128\over 15}\sqrt{\rho a^3\over\pi}\right]  \label{re.19c}
\eeq
As a consistency check, we see that it reproduces the pressure (\ref{re.19})
using the standard definition $P = \rho^2 \del({\cal E}/\rho)/\del\rho$.

\section{Effective potential at finite temperature}

Ignoring the counter-terms, the effective potential in the one-loop approximation and at 
finite temperature was obtained in (\ref{fluct.8b}) as function of the condensate $v$.
The excitation energy $\omega$ is now given by the general formula (\ref{fluct.7}).
Taking the derivative with respect to $v$, we then find that the minimum of $U_{eff}$
is shifted to
\beq
      v^2 = {\mu\over\lambda} - \!\int\!{d^3k\over(2\pi)^3}{4\ve - 4\mu 
          + 6\lambda v^2\over\omega}
      \left[{1\over 2} + {1\over e^{\beta\omega} - 1}\right]                \label{temp.3}
\eeq
due to thermal fluctuations. Here, $\mu$ is a function of temperature for fixed 
total particle density $\rho$. The above equation then gives the condensate 
as an implicit function of temperature. The contribution from thermal
fluctuations is finite. Therefore, the divergences are removed by 
the counter-terms already introduced at zero temperature.

With thermal fluctuations present, $v^2$ may differ considerably from $\mu/\lambda$. From (\ref{fluct.7}) we see that the excitation energy no longer 
satisfies $\omega \propto k$ in the long-wavelength limit. In other words, 
at finite temperature when the classical field 
takes on a modified value, the Goldstone theorem seems to be violated in the one-loop 
approximation considered up to now.

It is easy to see how the effective potential can be improved so that 
the Goldstone theorem is restored. At
finite temperature there must be additional effects taken into consideration
which changes the thermodynamic chemical potential $\mu$ into an effective
chemical potential $\td{\mu}$ so that the condensate is
again given by $v^2 = \td{\mu}/\lambda$. This will give a linear dispersion relation at low energy. The minimum of the effective potential will then be at
\beq
       v^2 = v_0^2 
           - \int\!{d^3k\over(2\pi)^3}{4\ve + 2\td{\mu}\over\omega}
               {1\over e^{\beta\omega} - 1}                           \label{temp.4}
\eeq
where $v_0^2$ is the zero-temperature result (\ref{re.10}). This is now consistent with
the Bogoliubov dispersion relation 
\beq
     \omega = \sqrt{\ve(\ve + 2\td{\mu})}                      \label{temp.5}
\eeq
which will be verified in the following where the ring-improved effective potential is 
derived.

The value of the condensate decreases with increasing temperatures and 
becomes zero when $\td{\mu} = 0$. This defines the critical temperature for 
the system. At higher temperatures it is in the phase of unbroken symmetry
where the dispersion relation simplifies to $\omega = \ve - \td{\mu}$ as it is for a
system of free particles.

\subsection{Ring corrections to the effective potential}

The chemical potential $\mu$ represents the energy of a adding or removing a 
single particle from the system of interacting bosons. It represents the
self energy of the complex field $\psi$. We see from (\ref{fluct.4}) that in the
symmetry-broken phase each of the two real modes of the field has self energies
equal to respectively $\mu - 3\lambda v^2$ and $\mu - \lambda v^2$ 
within the Gaussian approximation. Including the interactions in the full
Lagrangian (\ref{fluct.2}) these values will be modified by radiative loop 
corrections. When the system is at a non-zero temperature, the additional
contributions to the self energies will depend upon temperature. This will
then enable us to define a new, temperature-dependent chemical potential
$\td{\mu}$.

In the Gaussian approximation used in the previous section we found 
that the one-loop contribution to the effective potential (\ref{fluct.6})
followed directly from the inverse of the field propagator 
$D_{ab} = \ex{\chi_a\chi_b}$. After Fourier transformation we know from 
(\ref{fluct.4}) that it has the matrix form
\beq
     D_{ab}^{-1} = \left(\begin{array}{cc} e_{1\bk}  &  -\omega_n  \\
              \omega_n  & e_{2\bk}  \end{array}\right)    	   \label{ring.1}
\eeq
where $e_{1\bk} = \ve_\bk - \mu + 3\lambda v^2$ and $e_{2\bk} = \ve_\bk - \mu + \lambda v^2$.                 This lowest correction to the classical result is thus
\beq
     \Omega_1 = {1\over 2\beta}\Tr\ln D^{-1}
              = {1\over 2\beta}\sum_{n,\bk}\ln(\omega_n^2 + e_{1\bk}e_{2\bk}) 
                                                                \label{ring.4}
\eeq
which gave the standard expression (\ref{fluct.8}).

The matrix (\ref{ring.1}) is the inverse of the free propagator which is used
in the simplest version of the one-loop approximation. It was pointed out a
long time ago in connection with the effective potential for scalar, 
relativistic theories \cite{DJ,Weinberg} that self energy corrections
to the field propagator gave important contributions to the free energy at
finite temperatures. More recently these daisy corrections have been investigated
in more detail \cite{Carrington} and are now often sometimes called ring corrections
to the effective potential. We will here see that they also play a crucial
role in the non-relativistic theory in saving 
the Goldstone theorem at non-zero temperatures.
A technique which incorporates the use of resumed propagators in a self-consistent way is the effective action for composite operators \cite{CJT}. In the Appendix we show that this method leads to the same result as found here, to the order considered.
\begin{figure}[htb]
\begin{center}
\mbox{\psfig{figure=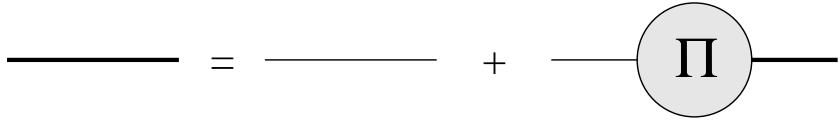,width=11cm,angle=0,height=1.5cm}}
\end{center}
\caption[A graphical representation of the Schwinger-Dyson equation. The thin and thick lines represent the free and interacting propagators, respectively.]{\footnotesize A graphical representation of the Schwinger-Dyson equation (\ref{ring.5a}). The thin and thick lines represent the free and interacting propagators, respectively.}
\label{bec3D-fig:sd}
\end{figure}

Denoting the interacting propagator by $\bar{D}_{ab}$, it will in
general satisfy the Dyson-Schwinger equation (\ref{prop.5}). If $\Pi_{ab}$ 
is the full, one-particle irreducible self energy, the equation then becomes
\beq										   
	 \bar{D}_{ab} =	 D_{ab} + D_{ac}\Pi_{cd}\bar{D}_{db}  \label{ring.5a}
\eeq
as shown in Fig. 2. Using now this propagator in
the one-loop approximation, we get the modified contribution
\beq
     \bar{\Omega}_1  = {1\over 2\beta}\Tr\ln \bar{D}^{-1}    \label{ring.5b}
\eeq
to the effective potential. Since we can write (\ref{ring.5a}) in the form
\beq
     \bar{D}_{ab}^{-1} = D_{ab}^{-1} - \Pi_{ab}                \label{ring.6}
\eeq
we have 
\beq
     \bar{\Omega}_1 & = & {1\over 2\beta}\Tr\ln D^{-1}(1 - D\Pi) \nn \\
                    & = & {\Omega}_1 - {1\over 2\beta}\Tr\left[D\Pi 
   + {1\over 2}(D\Pi)^2 + {1\over 3}(D\Pi)^3 + \cdots \right]  \label{ring.7}
\eeq
These additional contributions to the lowest order result are called ring 
corrections as seen in Fig. 3. 
\begin{figure}[htb]
\begin{center}
\mbox{\psfig{figure=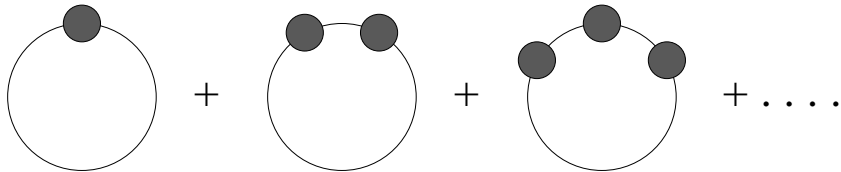,width=11cm,angle=0,height=2.2cm}}
\end{center}
\caption[A sketch of the lowest order ring diagrams.]{\footnotesize A sketch of the lowest order ring diagrams.}
\label{bec3D-fig:ring}
\end{figure}

Perturbative calculations will in the following show that the off-diagonal
self energies $\Pi_{12} = -\Pi_{21}$ vanish for zero external energy. 
This follows in general from time-reversal invariance, which implies that
the off-diagonal self energies must be odd in the Matsubara frequency 
$\omega_n$.
The structure of the interacting propagator will
thus be the same as for the free propagator (\ref{ring.1}) but with the
diagonal elements changed to
\beq
    \bar{e}_{1\bk} = \ve_\bk - \mu + 3\lambda v^2 - \Pi_{11} \label{ring.8}
\eeq
and
\beq
    \bar{e}_{2\bk} = \ve_\bk - \mu + \lambda v^2 - \Pi_{22}  \label{ring.9}
\eeq
The ring-corrected one-loop contribution to the effective effective potential 
(\ref{ring.5b}) is thus
\beq
    \bar{\Omega}_1 = {1\over 2\beta}\sum_{n,\bk}
    \ln(\omega_n^2 + \bar{e}_{1\bk}\bar{e}_{2\bk})              \label{ring.10}
\eeq
and will be of exactly the same form as the lowest order result (\ref{fluct.8})
except for the modified dispersion relation
\beq
     \bar{\omega}_\bk = \sqrt{(\ve_\bk - \mu + 3\lambda v^2 - \Pi_{11})
     (\ve_\bk - \mu + \lambda v^2 - \Pi_{22})}            \label{ring.11}
\eeq                                                    
for the elementary excitations above the ground state. 

The Goldstone theorem can now be satisfied also at non-zero temperatures if 
the chemical potential is related to the value of the field $ v$ in the 
minimum of the effective potential by
\beq
      \mu = \lambda v^2 - \Pi_{22}                         \label{ring.12}
\eeq
This is our form of the Pines-Hugenholtz relation  \cite{PH} which is usually written
is the complex field basis. A simple derivation of this theorem is given in \cite{SB}. Introducing the effective chemical potential
\beq
      \td{\mu} = \mu + \Pi_{22}                                \label{ring.13}
\eeq
the expectation value of the field when
the system is in thermal equilibrium is then given by 
$ v = (\td{\mu}/\lambda)^{1/2}$ as it is at zero temperature. This will be
demonstrated in the next section.

The dispersion relation (\ref{ring.11}) now becomes
\beq
     \bar{\omega}_\bk = \sqrt{\ve_\bk(\ve_\bk + 2\td{\mu} + \Pi_{22} - \Pi_{11})} 
                                                               \label{ring.14}
\eeq                                                    
and is by construction linear in the momentum $k$ in the long-wavelength 
limit. When the self energies are calculated in the following to the lowest 
order in perturbation theory, the difference $\Pi_{22} - \Pi_{11}$ will be 
negligible. The dispersion relation at finite temperatures is thus the same as at 
zero temperature (\ref{fluct.9}) when the chemical potential $\mu$ is replaced by 
$\td{\mu}$, i.e. it has the desired form (\ref{temp.5}). Similarly,
the inverse full propagator (\ref{ring.6}) has the form
\beq
     \bar{D}_{ab}^{-1} = \left(\begin{array}{cc} \ve_\bk + 2\td{\mu} &  -\omega_n  \\
              \omega_n  & \ve_\bk  \end{array}\right)    	   \label{ring.15}
\eeq
if the Goldstone theorem is to be satisfied.

\begin{figure}[h]
\begin{center}
\mbox{\psfig{figure=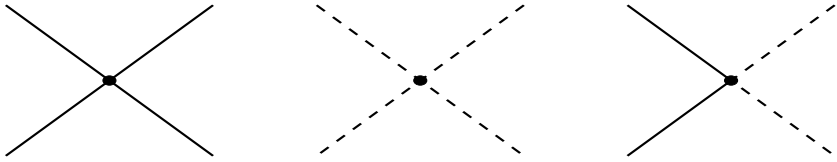,width=11cm,angle=0,height=2cm}}
\end{center}
\caption[Four-point vertices. Solid lines represent the $D_{11}$-propagator and dashed lines the $D_{22}$-propagator.]{\footnotesize Four-point vertices. Solid lines represent the $D_{11}$-propagator and dashed lines the $D_{22}$-propagator.}
\label{bec3D-fig:4ver}
\end{figure}

\subsection{One-loop contributions to the self energies}

The interacting propagators and the corresponding self energies can be obtained from
the full partition function $\Xi$ of the system. It is given by the functional integral
(\ref{fluct.5}). In the exponent we see that there is the usual coupling $(-\lambda)$ 
between four excitations $\chi$ as shown in Fig.4 but also a new coupling of magnitude 
$(-\lambda v)$ between three excitations due to the presence of the condensate 
$\ex{\psi} = v/\sqrt{2}$. Since we have chosen this expectation value to be real, this latter 
coupling will generate only the two vertices shown in Fig.5.

\begin{figure}[htb]
\begin{center}
\mbox{\psfig{figure=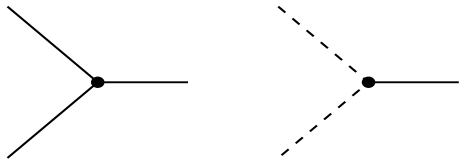,width=6cm,angle=0,height=2cm}}
\end{center}
\caption[Three-point vertices due to the non-zero condensate.]{\footnotesize Three-point vertices due to the non-zero condensate.}
\label{bec3D-fig:3ver}
\end{figure}

Separating out the free partition function $\Xi_0$, we will here calculate the part $\Xi_1 = 
\Xi/\Xi_0$ due to the interactions in lowest order of perturbation theory, i.e. include terms to
order $\lambda$. Since the chemical potential is now defined to be $\td{\mu} = \lambda v^2$,
we must include all diagrams where the four-excitation coupling occurs once or the three-excitation
coupling occurs twice. Only connected diagrams will contribute to $\ln{\Xi_1}$ and they are given
with the corresponding combinatorial factors in Figure 6. The lines in the diagrams represent
here the free propagator $D_{ab}$ given in (\ref{ring.1}).

\begin{figure}[htb]
\begin{center}
\mbox{\psfig{figure=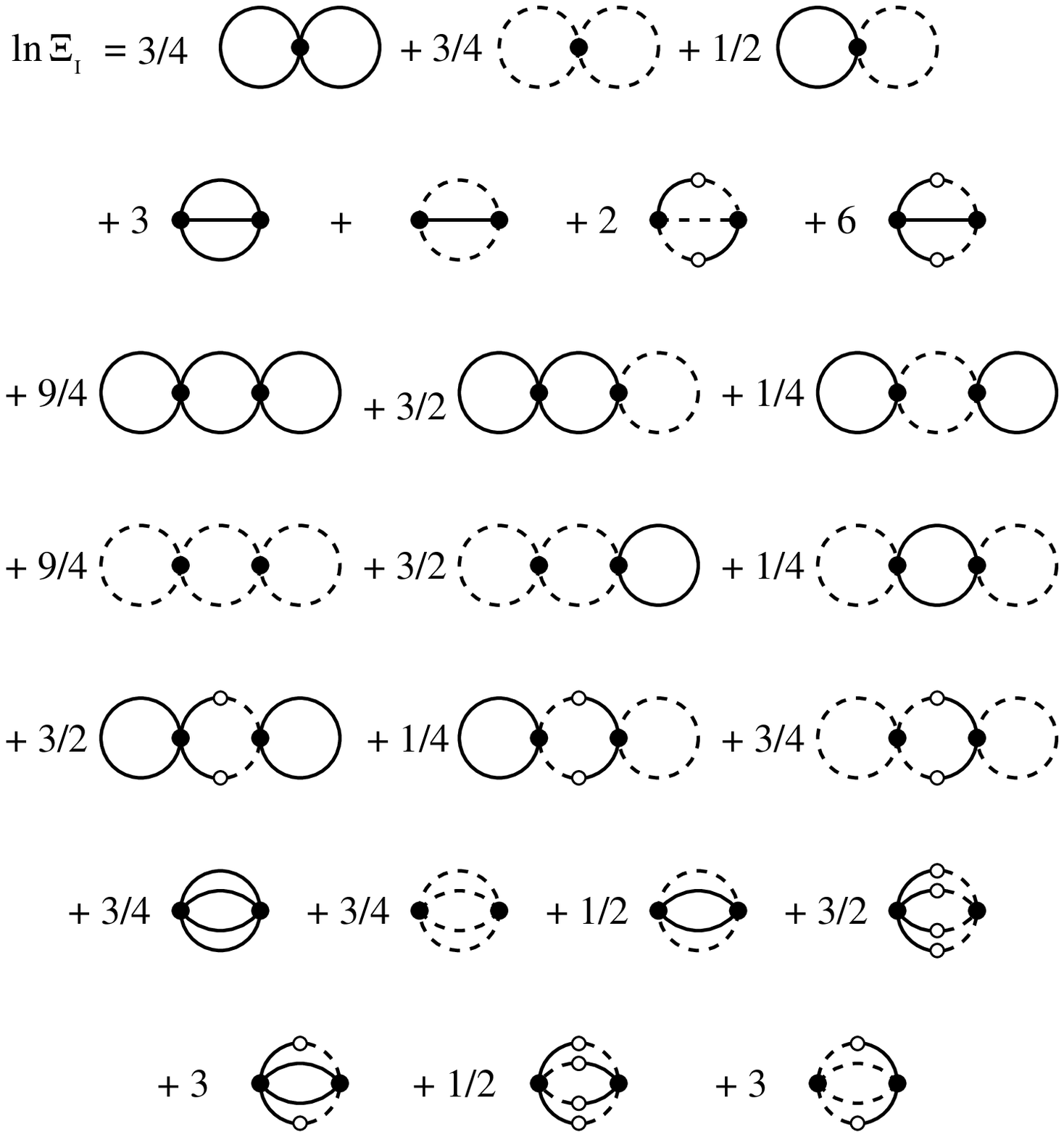,width=14cm,angle=0,height=12cm}}
\end{center}
\caption[Two- and three-loop diagrams contributing to $\ln\Xi_I$. Only two-loop diagrams are considered here.]{\footnotesize Two- and three-loop diagrams contributing to $\ln\Xi_I$. Only two-loop diagrams are considered here.
The lines represent the free propagator $D_{ab}$.}
\label{bec3D-fig:lnxi}
\end{figure}

We can now derive the full propagator $\bar{D}_{ab} = \ex{\chi_a\chi_b}$ from the partition function 
$\Xi$  by taking the derivative with respect to the free propagator $D_{ab}$ as one can see from 
the functional integral (\ref{fluct.5}). One then finds
\beq
    \bar{D}_{ab} = -2\frac{\delta\ln\Xi}{\delta D^{-1}_{ab}} = 
                 - 2\frac{\delta\ln\Xi_0}{\delta D^{-1}_{ab}} 
                 + 2D_{ca}D_{bd}\frac{\delta\ln\Xi_1}{\delta D_{cd}}        \label{self.1}
\eeq
The first term is just the free propagator while the last term gives the self energy. In terms of 
diagrams, the derivative is obtained directly from the bubble diagrams
in Fig. 6 by opening up the corresponding lines in all of the loops. For example, $\bar{D}_{11}$ 
is given by the diagrams shown in Fig. 7. 

\begin{figure}[htb]
\begin{center}
\mbox{\psfig{figure=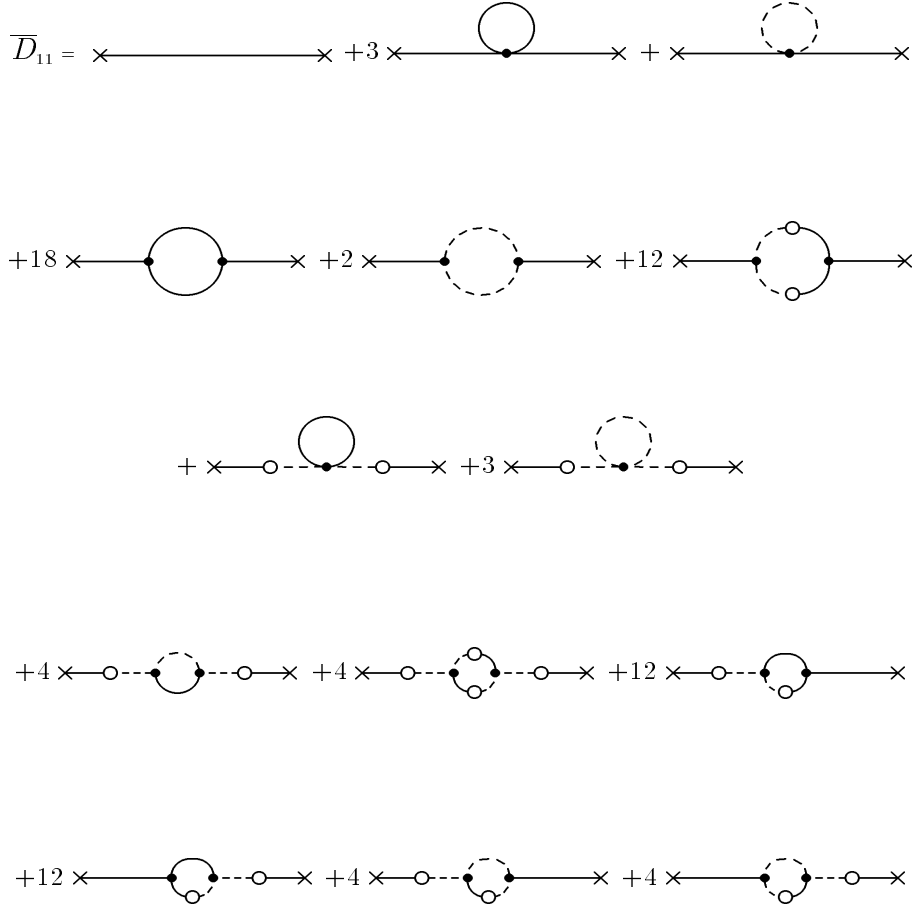,width=12cm,angle=0,height=12cm}}
\end{center}
\caption[The lowest order diagrams contributing to the interacting propagator $\td{D}_{11}$. Crosses indicate that the external propagators should be included.]{\footnotesize The lowest order diagrams contributing to the interacting propagator $\td{D}_{11}$. Crosses indicate that the external propagators should be included. The lines represent the free propagator 
$D_{ab}$.}
\label{bec3D-fig:D11}
\end{figure}

\begin{figure}[hbt]
\begin{center}
\mbox{\psfig{figure=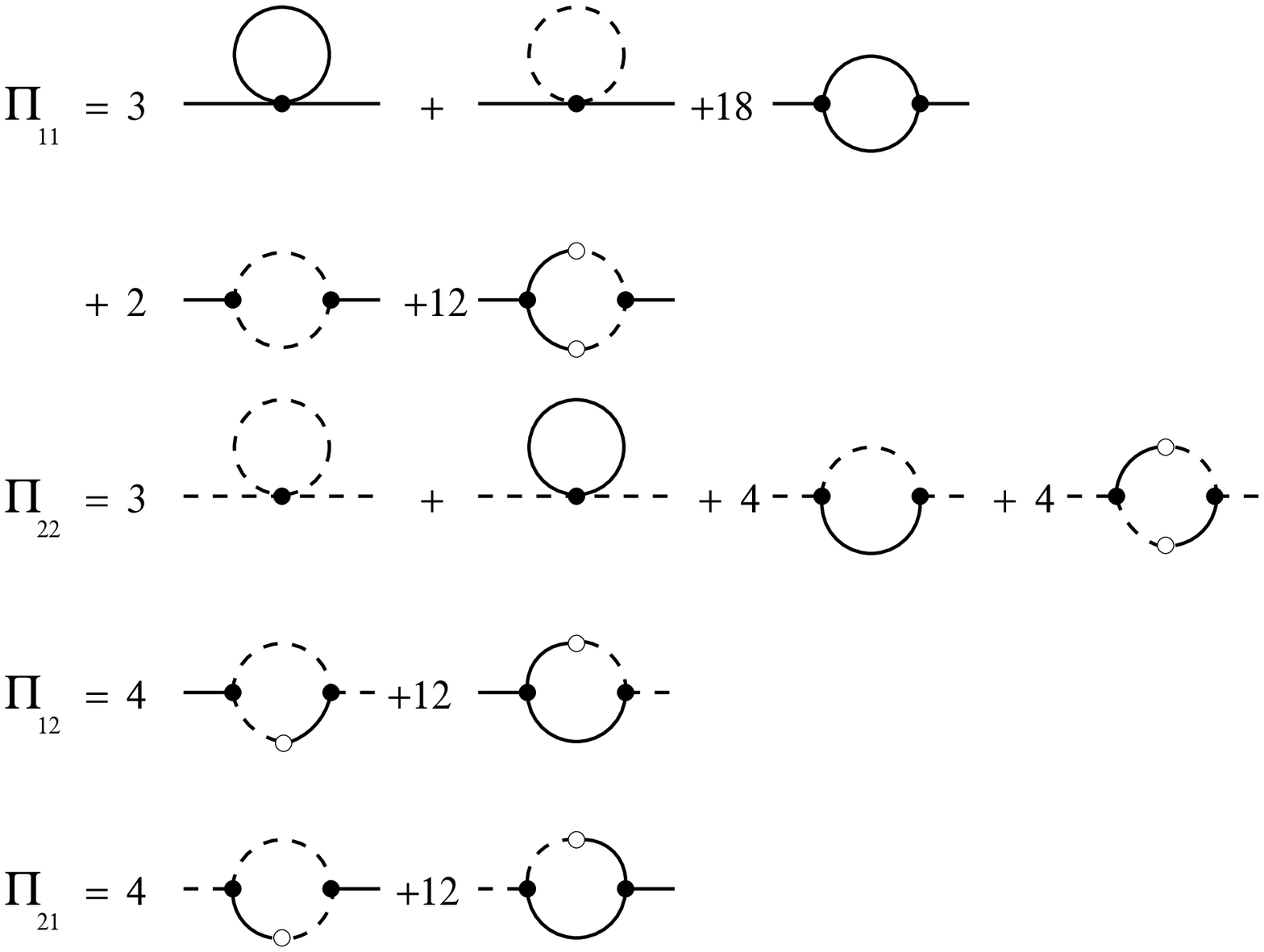,width=11cm,angle=0,height=8cm}}
\end{center}
\caption[Self energies to lowest order in the loop expansion. The external propagators should not be included.]{\footnotesize Self energies to lowest order in the loop expansion. The external propagators should not be included.}
\label{bec3D-fig:Pi}
\end{figure}

By cutting off the external lines in these diagrams,
we can read off the different self energies $\Pi_{cd}$ shown in Fig. 8. This is equivalent to using
(\ref{ring.5a}) in lowest order of perturbation theory where it gives $\bar{D}_{ab} = D_{ab} + 
D_{ac}\Pi_{cd}D_{db}$ from which $\Pi_{cd}$ can be isolated.

An infinite set of higher loop contributions to the free energy can now be obtained by replacing
the free propagator $D_{ab}$ in the diagrams in Fig. 6 with the full propagator $\bar{D}_{ab}$. This
iteration then generates diagrams which correspond to the ``super daisies'' of Dolan and Jackiw
\cite{DJ}. They have more recently been investigated in more detail by Carrington
\cite{Carrington} and others \cite{AE}. In this improved
perturbation theory the self energies in Fig. 8 are then to be calculated with the full propagators.
In order for the Goldstone theorem to be satisfied, we know that $\bar{D}_{ab}$ must have the
form (\ref{ring.15}) involving the effective chemical potential $\td{\mu}$. If this
approximation method is consistent, the unknown function $\td{\mu}(T)$ can then be determined
from the Pines-Hugenholtz relation (\ref{ring.13}) which now will constitute a gap equation.

There are other diagrams in addition to those given in Fig. 6 and Fig. 7, but they all vanish. Tadpole 
diagrams, such as 
$
                \plot 10 3.2 20 3.2 /  
  		\plot 10 3.0 20 3.0 / 		
		\plot 10 3.2 20 3.2 /  
		\plot 10 3.4 20 3.4 /  
		\plot 10 3.6 20 3.6 /  
		\plot 10 3.8 20 3.8 / 
        	\thicklines
        	\put {\circle{12}} [B1] at 32 3.3 
$
\hspace{1.5cm},
are zero because $\ex{\chi_1} = 0 = \ex{\chi_2}$. Diagrams with a $\bar{D}_{12}$-loop 
$
                \plot 10 3.2 15 3.2 /  
  		\plot 10 3.0 15 3.0 / 		
		\plot 10 3.2 15 3.2 /  
		\plot 10 3.4 15 3.4 /  
		\plot 10 3.6 15 3.6 /  
		\plot 10 3.8 15 3.8 / 
                \put {\circle{4}} [B1] at 28.5 3.3 
        	\thicklines
\circulararc 156 degrees from 26.5 5.7 center at 21 3.3
\circulararc 156 degrees from 26.4 5.7 center at 21 3.3
\circulararc 156 degrees from 26.3 5.7 center at 21 3.3
                \setdashes <1.50pt>
\circulararc 156 degrees from 15 3.3 center at 21 3.3
                \setdashes <1.50pt>
\circulararc 156 degrees from 15.1 3.3 center at 21 3.3
                \setdashes <1.50pt>
\circulararc 156 degrees from 15.2 3.3 center at 21 3.3
$
\hspace{1.0cm}
are odd in the Matsubara frequency and thus vanish upon summation. 
The perturbative contributions to the diagonal self energies are non-zero. They depend in general
on external momentum and energy. We have assumed a spatially constant condensate and can thus take the
external momenta in the self energies to be zero. For the external energies we will also take these
to be zero. This is obviously an approximation. At sufficiently high temperature, the zero-energy 
Matsubara modes should dominate the partition function and the approximation should then
be reasonable. Especially for $\Pi_{22}$ which we will need in the Pines-Hugenholtz relation 
(\ref{ring.13}), the approximation should be good because the pole in the corresponding real-time 
propagator is at zero energy.
The off-diagonal self energies $\Pi_{12}$ and $\Pi_{21}$ in Fig. 8 both vanish
at zero external energy. As earlier 
stated, this follows to all orders in perturbation theory from time-reversal invariance. 

From the diagrams in Fig. 8 we then find the self energy
\beq
    \Pi_{11} & = & {1\over\beta}\sum_n\int{d^3k\over(2\pi)^3}
                   \left[3(-\lambda)\bar{D}_{11}(k)  + (-\lambda)\bar{D}_{22}(k)
     + 18(-\lambda v)^2\bar{D}_{11}(k)\bar{D}_{11}(-k) \right. \nn \\
             & + & \left.2(-\lambda v)^2 \bar{D}_{22}(k)\bar{D}_{22}(-k)+ 12(-\lambda v)^2 \bar{D}_{12}(k)\bar{D}_{12}(-k)\right]
\eeq
where we use $k \equiv (\omega_n,\bk)$.
With the full propagators from (\ref{ring.15}) this can be simplified to
\beq
    \Pi_{11} = - {\lambda\over\beta}\sum_n\int{d^3k\over(2\pi)^3}
                \left[{4\ve + 2\td{\mu}\over\omega^2 + \omega_n^2} 
              - {2\lambda v^2(10\ve^2 + 4\ve\td{\mu}+ 4\td{\mu}^2 - 6\omega_n^2)
                \over(\omega^2 + \omega_n^2)^2}\right]
                                                                \label{pi.4}
\eeq
where $\omega^2 = \ve(\ve + 2\td{\mu})$. The Matsubara summations are now easy to do. From the 
standard sum
\beq
     \sum_{n=-\infty}^\infty {1\over \omega^2 + \omega_n^2} 
   = {\beta\over\omega}\left[{1\over 2} + {1\over e^{\beta\omega} - 1}\right]
\eeq
follows directly by differentiation
\beq
     \sum_{n=-\infty}^\infty {\omega^2\over (\omega^2 + \omega_n^2)^2} 
   = {\beta\over 2\omega}\left[{1\over 2} + {1\over e^{\beta\omega} - 1} 
   + \beta\omega{e^{\beta\omega}\over \left(e^{\beta\omega} - 1\right)^2}\right]
\eeq
With $\omega_n^2$ instead of $\omega^2$ in the numerator on the left-hand side, we get the same
result except for an opposite sign in front of the last term. The zero-temperature self energy
is then found to be
\beq
     \Pi_{11}(T=0) &=& -\lambda\int{d^3k\over(2\pi)^3}\left[{2\ve + 4\td{\mu}\over\omega}
                   - {\td{\mu}(5\ve^2 + 2\ve\td{\mu} + 2\td{\mu}^2)\over\omega^3}\right]\nonumber\\
&=&-\lambda\int{d^3k\over(2\pi)^3}\left[{2\ve + \td{\mu}\over\omega}
                   - {2\td{\mu}(\ve - \td{\mu})^2\over\omega^3}\right]
                                                                \label{pi.5}
\eeq
In the last line we have regrouped the terms to show the difference between 
$\Pi_{11}$ and $\Pi_{22}$ found below.
The leading divergences will be cancelled by the counter-terms. This will explicitly be
demonstrated in the case $\Pi_{22}$ which we evaluate next. From the diagrams in Fig. 8
it is given as
\beq
     \Pi_{22}& = & {1\over\beta}\sum_n\int{d^3k\over(2\pi)^3}
                   \left[3(-\lambda)\bar{D}_{22}(k) + 
                   (-\lambda)\bar{D}_{11}(k)\right.\nn \\
             & + & \left.4(-\lambda v)^2\bar{D}_{11}(k)\bar{D}_{22}(-k)
               + 4(-\lambda v)^2 \bar{D}_{12}(k)\bar{D}_{12}(-k)\right] \nn \\
             & = & - {\lambda\over\beta}\sum_n\int{d^3k\over(2\pi)^3}
                 \left[{4\ve + 6\td{\mu}\over\omega^2 + \omega_n^2} 
               - {4\td{\mu}(\ve^2 + 2\ve\td{\mu} + \omega_n^2)
                 \over(\omega^2 + \omega_n^2)^2}\right]\nonumber\\
             & = & - {\lambda\over\beta}\sum_n\int{d^3k\over(2\pi)^3}
                     {4\ve + 2\td{\mu}\over\omega^2 + \omega_n^2}
\eeq
A summation over Matsubara frequencies gives at non-zero temperature
\beq
     \Pi_{22}(T) = -\lambda\int{d^3k\over(2\pi)^3}{4\ve + 2\td{\mu}\over\omega}
                    \left[{1\over 2} + {1\over e^{\beta\omega} - 1}\right]         \label{pi.8}
\eeq
The first part of the integral which represents the zero-temperature self energy, is 
ultraviolet divergent. However, from (\ref{ct.1}) we see that we should add in the 
counter-terms $\delta\mu - v^2\delta\lambda$. Taking these from (\ref{re.4}) and 
(\ref{re.5}), with the replacement $\mu\rightarrow\td{\mu}$, we obtain the renormalized and finite result
\beq
     \Pi_{22}(T)  = - \lambda\int{d^3k\over(2\pi)^3}\left[{2\ve + \td{\mu}\over\omega}
                    + {\td{\mu}\over\ve} - 2 + {4\ve + 2\td{\mu}\over\omega}
                     {1\over e^{\beta\omega} - 1}\right]
                                                                \label{pi.10}
\eeq
Similarly, an expansion for high momenta of the zero-temperature part of $\Pi_{11}$ in (\ref{pi.5}) shows that the counter-terms $\delta\mu - 3v^2\delta\lambda$ introduced in (\ref{ct.1}), again with $\mu\rightarrow\td{\mu}$, exactly cancel the ultraviolet divergences in $\Pi_{11}$. 

The assumption that the Goldstone theorem is fulfilled, i.e. the Pines-Hugenholtz relation (\ref{ring.12}), can now be shown to equivalent to the minimalization criterion (\ref{temp.4}) for a one-loop potential calculated with the full propagator $\bar{D}_{ab}$. 
Combining (\ref{re.10}) with $\mu \rightarrow \td{\mu}$ in the one-loop term and (\ref{temp.4}) we find 
\beq
v^2 &=& \frac{\mu}{\lambda} - \int{d^3k\over(2\pi)^3}\left[{2\ve + \td{\mu}\over\omega} + {\td{\mu}\over\ve} - 2 + {4\ve + 2\td{\mu}\over\omega}
{1\over e^{\beta\omega} - 1}\right]\nonumber\\
&=& \frac{\mu}{\lambda} + \frac{1}{\lambda}\Pi_{22}(T)
                                                                \label{pi.11}
\eeq
when making use of the result (\ref{pi.10}).
This is the same as equation (\ref{ring.12}) which embodies the Goldstone theorem .

As mentioned at the beginning of this section, the self energies are calculated
using the dispersion relation $\omega = \sqrt{\ve(\ve + 2\td{\mu})}$, i.e. neglecting terms of ${\cal O}(\Pi_{11} - \Pi_{22})$. Using the renormalized self energies, we see that the difference at zero temperature is
\beq
\Pi_{11} - \Pi_{22} = 2\lambda\mu\int{d^3k\over(2\pi)^3}\left[{(\ve - \mu)^2\over\omega^3} - {1\over\ve}\right]
                                                                 \label{pi.12}
\eeq
The integral is finite at high momenta due to the counter-term, but is infrared divergent. This is caused by the exchange diagram in $\Pi_{11}$ with two $\bar{D}_{22}$-lines
which diverges as external energy and momenta are taken to zero. 
The infrared divergence signals the onset of long-distance 
effects which are not properly handled by the present one-loop approximation.
As discussed by Kapusta \cite{Kapusta_2}, one can cure the divergence using a non-zero external
energy. However, this is not a problem here. At low temperature, the self energies are of ${\cal O}(\lambda\mu^{3/2})$ and will be neglected all together, being small compared to $\mu$. On the other hand, at high temperature, i.e. near the critical temperature $T_C$, $\Pi_{22}$ is comparable to $\mu$ and the self energies must be included. In this regime we can use the high-temperature limits of $\Pi_{11}$ and $\Pi_{22}$ which below are shown to be equal to leading order. Thus, in the temperature range where it should be considered, $\Pi_{11} - \Pi_{22} = 0$.

At high temperatures $T\gg\td{\mu}$ the exchange diagrams may be neglected. The dominating terms in the remaining diagrams go like $\ve/(\omega(e^{\beta\omega}-1))$, and the combinatorial prefactors are the same for the two self energies. Expanding (\ref{pi.4}) and (\ref{pi.10}) in powers of $\td{\mu}/T$ we find to leading order
\beq
\Pi_{22}(T) = \Pi_{11}(T) = -4\lambda\zeta{(3/2)}\left({mT\over 2\pi}\right)^{3/2}
                                                                                       \label{pi.13}
\eeq
This result will be found if one uses free propagators $D_{ab}$ in the self energy loops since $\td{\mu}$ can be neglected to leading order.

In the opposite limit  $T\ll\td{\mu}$, we can take $\td{\mu}$ approximately equal to $\mu$. Diagrams with a $\bar{D}_{11}$-loop are suppressed by a factor
$T^2/\mu^2$ compared to the others. We then obtain for the leading terms of the self energy
\beq
\Pi_{22}(T) = -\frac{10\lambda}{3\pi^2}(m\mu)^{3/2} - \frac{\lambda T^2m^{3/2}}{6\mu^{1/2}}
                                                                 \label{pi.14}
\eeq
The first part is seen to be of the order ${\cal O}(\sqrt{\rho a^3})$ smaller than the chemical potential and
the second term even smaller by a factor $(T/\td{\mu})^2$. To leading order, we can thus neglect the self energies at these low temperatures and set $\td{\mu} = \mu$. Thus, the effects of ring corrections on the
thermodynamics of the system will only appear at high temperature.

\subsection{Condensate and pressure at finite temperature}
Including the effects of ring corrections, we can now write down the renormalized finite temperature effective potential
\beq
U_{eff}(v,\mu ,T) &=& -\frac{\mu}{2}v^2 + \frac{\lambda}{4}v^4 + {1\over 2}\int\!{d^3k\over(2\pi)^3}\!\left(\omega - \ve + \td{\mu} - 2\lambda v^2 + {\lambda^2v^4\over 2\ve}\right) \nonumber\\
&+& T \int\!{d^3k\over(2\pi)^3}\!\ln(1 - e^{-\beta\omega})
                                                                \label{eff.1}
\eeq
From (\ref{pi.11}) we know that it has a minimum at $v = (\td{\mu}/\lambda)^{1/2}$. In the minimum we then have 
the pressure
\beq
P(\mu ,T) &=& -U_{eff}(\mu ,T) = \frac{1}{4\lambda}(\mu^2 - \Pi_{22}^2) - {1\over 2}\int\!{d^3k\over(2\pi)^3}\!\left[\omega - \ve - \td{\mu} + \frac{\td{\mu}^2}{2\ve}\right]\nonumber\\
&-& T\int\!{d^3k\over(2\pi)^3}\!\ln(1 - e^{-\beta\omega})
                                                                \label{eff.2}
\eeq
The total density of particles is again found by using $\rho = \partial P/\partial\mu$ which gives
\beq
\rho &=& {\mu\over 2\lambda} - {1\over 2}\int\!{d^3k\over(2\pi)^3}\!
          \left[{\ve\over\omega} + {\td{\mu}\over\ve} - 1 + \frac{2\ve}{\omega(e^{\beta\omega}-1)}\right] \nonumber\\
&=& {\td{\mu}\over 2\lambda} + {1\over 2}\int\!{d^3k\over(2\pi)^3}\!
\left[{\ve + \td{\mu}\over\omega} - 1 + \frac{2(\ve + \td{\mu})}{\omega(e^{\beta\omega}-1)}\right]
                                                                \label{eff.3}
\eeq
while the density of particles in the condensate is 
\beq
\rho_c = \frac{v^2}{2} = \frac{\td{\mu}}{2\lambda} = \frac{\mu}{2\lambda} -{1\over 2}\int\!{d^3k\over(2\pi)^3}\!\left[\frac{2\ve + \td{\mu}}{\omega} + \frac{\td{\mu}}{\ve} - 2 + \frac{4\ve + 2\td{\mu}}{\omega(e^{\beta\omega}-1)}\right]
                                                               \label{eff.4}
\eeq
The difference 
\beq
\rho_e = {1\over 2}\int\!{d^3k\over(2\pi)^3}\!\left[\frac{\ve + \td{\mu}}{\omega}\left(1 + \frac{2}{e^{\beta\omega}-1}\right) -1 \right]
                                                               \label{eff.5b}
\eeq
represents the density of particles in excited states with $\bk\neq 0$. Except for the replacement of $\mu$ with
$\td{\mu}$, this result agrees with what was obtained in \cite{BBD}. Now both the hard-core repulsion and 
thermal fluctuations cause the excitation of particles from the condensate. 
With increasing temperatures more and more particles are in excited states and 
at a critical temperature where $\td{\mu} = 0$, the condensate becomes zero.  We then have a phase transition
to the normal phase. In the following we will see that this critical temperature is the same as for a free Bose 
gas to the accuracy we are working.

Equation (\ref{eff.3}) gives upon inversion the chemical potential as a function of the particle density. The first integral on the right-hand side was done in Section 3.4, and we can now simply replace $\mu$ with $\td{\mu}$ in Eq. (\ref{re.14}). The second integral 
\beq
     I_T \equiv \int\!{d^3k\over(2\pi)^3}\frac{\ve}{\omega(e^{\beta\omega}-1)} 
                                                               \label{eff.6}
\eeq
must be done numerically. However, in the two important temperature ranges we can easily find good analytical
approximations.

At low temperature we again take $\td{\mu}\simeq\mu$. The dominant contributions to the integral comes from 
small values of $\ve$ and we can thus take $\ve + 2\mu\simeq 2\mu$ in the denominator. Then we have the result
\beq
I_T = \frac{\pi^2m^{3/2}T^4}{60\mu^{5/2}}
                                                               \label{eff.7}
\eeq
Since the loop corrections are small, we can set $\mu = 2\lambda\rho$ on the right hand side of (\ref{eff.3}). 
Inversion then trivially gives
\beq
\mu = 2\lambda\rho + \frac{8\lambda}{3\pi^2}(2m\lambda\rho)^{3/2} + \frac{\lambda\pi^2 m^{3/2}T^4}{30(2\lambda\rho)^{5/2}} 
                                                               \label{eff.8}
\eeq
For comparison, the expression for the effective chemical potential takes the form
\beq
\td{\mu} = 2\lambda\rho_c = 2\lambda\rho - \frac{2\lambda}{3\pi^2}(2m\lambda\rho)^{3/2} - \frac{\lambda m^{3/2}T^2}{6\sqrt{2\lambda\rho}}
\eeq
Thus, at low $T$, the chemical potential {\em increases} with temperature, while the {\em effective} chemical potential, and thus the condensate density decreases with temperature. This temperature dependence in the same as obtained in the Bogoliubov approximation \cite{FW}.
The pressure now follows from Eq. (\ref{eff.2}). As already explained, we can neglect the contribution coming
from the self-energy. Thus
\beq
P = \frac{\mu^2}{4\lambda} - \frac{8m^{3/2}}{15\pi^2}\mu^{5/2} + P_T
                                                               \label{eff.9}
\eeq
where the temperature dependence is in the function
\beq
P_T = -T\int\!{d^3k\over(2\pi)^3}\!\ln(1 - e^{-\beta\omega}) = \frac{\pi^2 m^{3/2}T^4}{90\mu^{3/2}}
                                                               \label{eff.10}
\eeq
evaluated in the same approximation as above. The pressure at low temperatures is therefore
\beq
P(\rho) = \lambda\rho^2 + \frac{4m^{3/2}}{5\pi^2}(2\lambda\rho)^{5/2} + \frac{m^{3/2}\pi^2}{90}\frac{T^4}{(2\lambda\rho)^{3/2}}
                                                               \label{eff.11}
\eeq
At non-zero temperature we can obtain the energy density from the thermodynamic relation
\beq
{\cal E} = -\frac{\partial}{\partial\beta}(\beta P)_{\beta\mu}
\eeq
which now gives
\beq
{\cal E} = \frac{2\pi a}{m}\rho^2\left[1 + \frac{128}{15}\sqrt{\frac{a^3\rho}{\pi}}\,\right] + \frac{m^{3/2}}{240\pi}\frac{T^4}{\rho}\sqrt{\frac{\pi}{a^3\rho}}
                                                               \label{eff.12}
\eeq
It is seen that the effect of non-zero temperature comes in as a $T^4$ contribution as in the Stefan-Boltzmann
law for photons. In that case the excitations are mass-less because of gauge invariance, while here they are
Goldstone bosons resulting from a broken, continuous symmetry. 

In the opposite temperature range near the critical temperature $T_C$, we must include the effects of the
self energies. They have previously been obtained in (\ref{pi.13}) which gives for the effective chemical
potential (\ref{ring.13})
\beq
     \td{\mu} = \mu - 4\lambda\zeta{(3/2)}\left({mT\over 2\pi}\right)^{3/2} 
                                                                \label{eff.14}
\eeq
This quantity also enters the total density of particles in (\ref{eff.3}) where we now to leading order can
set $\td{\mu} = 0$. The temperature-dependent integral (\ref{eff.6}) then simplifies to
\beq
             I_T = \zeta{(3/2)}\left({mT\over 2\pi}\right)^{3/2} 
                                                               \label{eff.13}
\eeq
Since the chemical potential is now $\mu = 2\lambda(\rho + I_T)$, we have the relation
\beq
     \mu = 4\lambda\rho - \td{\mu}
\eeq
for temperatures just below the $T_C$. The critical temperature is defined by $\td{\mu} = 0$. Here the chemical potential takes the value $\mu_C = 4\lambda\rho$, and hence
(\ref{eff.14}) gives 
\beq
T_C = \frac{2\pi}{m}\left(\frac{\mu_C}{4\lambda\zeta{(3/2)}}\right)^{2/3} = \frac{2\pi}{m}\left(\frac{\rho}{\zeta{(3/2)}}\right)^{2/3}
                                                               \label{eff.17}
\eeq
which is just the textbook result \cite{Huang}. As seen in Fig. 9, the thermodynamic chemical potential $\mu$
increases continuously from zero to $\mu_C$ at the critical temperature.
On the other hand, the effective chemical potential $\td{\mu}$ decreases smoothly from zero temperature
and is by definition zero at $T_C$.

\begin{figure}[htb]
\begin{center}
\mbox{\psfig{figure=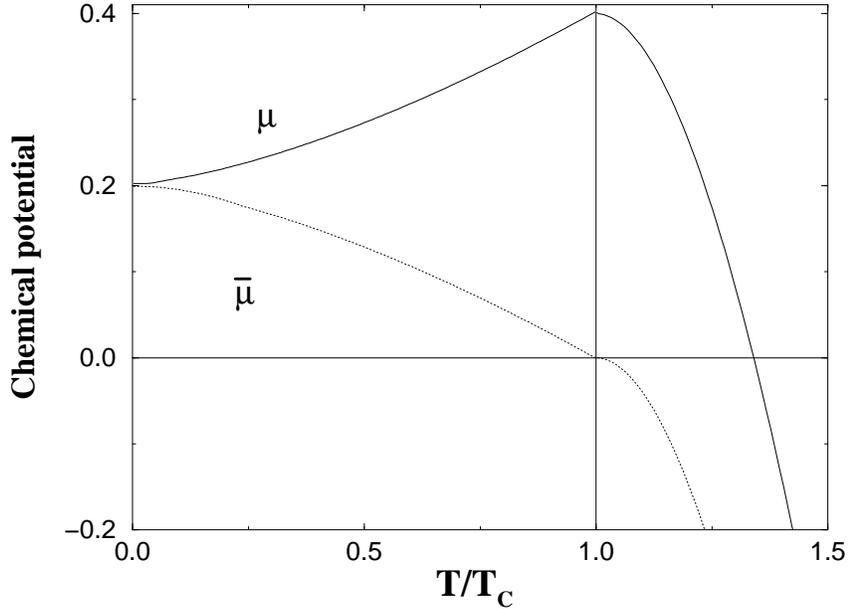,width=11cm,angle=270,height=8cm}}
\end{center}
\caption[The thermodynamic and effective chemical potentials, $\mu$ and $\td{\mu}$ plotted as functions of temperature for $ g=0.1$ and $m=1$.]{\footnotesize The thermodynamic and effective chemical potentials, $\mu$ and $\td{\mu}$ plotted as functions of temperature for $ g=0.1$ and $m=1$.
The results were obtained by matching the low and high temperature
expressions for the chemical potentials.}
\label{bec3D-fig:mu}
\end{figure}

In the general formula (\ref{eff.2}) we must now keep the contributions from the self energies when we
calculate the pressure for temperatures just below the critical temperature. Thus
\beq
P = \frac{1}{4\lambda}(\mu^2 - \Pi_{22}^2) + \left(\frac{m}{2\pi}\right)^{3/2}T^{5/2}\zeta(5/2) -
\left(\frac{mT}{2\pi}\right)^{3/2}\zeta(3/2)(\mu + \Pi_{22}) + {\cal O}(\td{\mu}^2)
\eeq
with $\Pi_{22}$ given in (\ref{pi.13}). In order to compare with results in the literature, we write the pressure as a function of the particle density
\beq
P = \frac{2\pi}{m}a\rho^2 - \left(\frac{m}{2\pi}\right)^2\zeta(3/2)^2aT^3 + \left(\frac{m}{2\pi}\right)^{3/2}\zeta(5/2)T^{5/2}
\eeq
Comparing with the result of Lee and Yang \cite{LeeYang_3}, we find that the second term has the wrong sign. The reason is the following: An expansion of the ring-corrected potential on powers of $\lambda$ shows that the two-loop contribution is counted twice. Higher order diagrams are reproduced correctly. We must correct for this double counting by subtracting the two-loop contribution, as given in perturbation theory, from the obtained result. This contribution is given by the three uppermost diagrams in Fig. 6. Near $T_C$ the propagators are equal to lowest order, $D_{11} = D_{22} = \ve/(\ve^2 + \omega_n^2)$. The total contribution is $-2(m/2\pi)^2\zeta(3/2)^2aT^3$. Subtracting this contribution, we thus have
\beq
P = \frac{2\pi}{m}a\rho^2 + \left(\frac{m}{2\pi}\right)^2\zeta(3/2)^2aT^3 + \left(\frac{m}{2\pi}\right)^{3/2}\zeta(5/2)T^{5/2}
                                                     \label{eff.20}
\eeq
which now agrees with \cite{LeeYang_3}. This result also comes out in the mean field approximation \cite{Huang,Anne}. The equation for the chemical potential is not modified by this correction at the present order of accuracy. In the same approximation we also find the the energy density
\beq
{\cal E}= \frac{2\pi}{m}a\rho^2 -\sqrt{\frac{m}{2\pi}}\zeta(3/2)^2a\rho T^{3/2} + 2(\frac{m}{2\pi})^2\zeta(3/2)^2aT^3 + \frac{3}{2}(\frac{m}{2\pi})^{3/2}\zeta(5/2)T^{5/2}
                                                               \label{eff.22}
\eeq
At the critical temperature these expressions reduce to
\beq
P(T_C)= \frac{2\pi}{m}\zeta(3/2)^{5/3}\zeta(5/2)\rho^{2/3} + \frac{4\pi}{m}a\rho^2
                                                               \label{eff.23}
\eeq
and
\beq
{\cal E}(T_C) = \frac{3\pi}{m}\zeta(3/2)^{5/3}\zeta(5/2)\rho^{2/3} + \frac{4\pi}{m}a\rho^2
                                                               \label{eff.24}
\eeq
which in the following are shown to equal the limits taken in the normal state, securing continuity at the critical temperature.

The special treatment at two-loop order could of course have been introduced at an earlier stage, but there would be no effects on the results found so far. When minimalizing the effective ring-corrected potential we discarded contributions from $\partial\Pi/\partial v$, being of ${\cal O}(\lambda^2)$. Since the contribution from differentiation of the two-loop potential is of the same order, this must be discarded as well, and the minimalization criterion remains unchanged. In the Appendix we briefly discuss how this is incorporated in the effective action for composite operators.

\subsection{Equation of state in the normal phase}
At temperatures above $T_C$, the condensate density is zero and the effective potential has its only minimum at $v=0$. The modified dispersion relation then reads ${\omega = \ve - \td{\mu}}$ with effective chemical potential $\td{\mu} = \mu +\Pi_{22} < 0$. Since $v=0$, the exchange diagrams vanish and the expressions for the self energies are 
\beq
\Pi_{11} = \Pi_{22} \equiv \Pi &=& {1\over\beta}\sum_n\int{d^3k\over(2\pi)^3} \left[3(-\lambda)\bar{D}_{22}(k) + (-\lambda)\bar{D}_{11}(k)\right] + \mbox{c.t.}\nonumber\\
&=& -\lambda\int\frac{d^3k}{(2\pi)^3}\frac{1}{e^{\beta\omega} - 1}
                                                               \label{eos.1}
\eeq
The Bose-integral is standard and the self energy becomes
\beq
\Pi = -4\lambda\left(\frac{mT}{2\pi}\right)^{3/2}\Li_{3/2}(e^{\beta\td{\mu}})
\eeq
We have here introduced the polylogarithmic function
\beq
\Li_n(x) = \sum_{k=1}^{\infty}\frac{x^k}{k^n}
                                                           \label{eos.2}
\eeq
with $\Li_n(1) = \zeta(n)$. 
Near $T_C$ we may approximate the exponential with unity and at $T_C$ the result (\ref{pi.13}) is reproduced. To the present accuracy we may set
$\td{\mu} = \mu$ in the self energy. With vanishing condensate the classical and zero point terms do not contribute to the pressure, which simplifies to
\beq
P &=& \left(\frac{m}{2\pi}\right)^{3/2}T^{5/2}\Li_{5/2}(\td{z}) + 2 \lambda \left(\frac{mT}{2\pi}\right)^3\Li^2_{3/2}(z)
                                                               \label{eos.3a}
\eeq
with $z = \exp(\beta\mu)$ and $\td{z} = \exp(\beta\td{\mu})$. To lowest order in $\lambda$ the pressure can be written
\beq
P = \left(\frac{m}{2\pi}\right)^{3/2}T^{5/2}\Li_{5/2}(z) - 2 \lambda\left(\frac{mT}{2\pi}\right)^3\Li^2_{3/2}(z)
                                                               \label{eos.3b}
\eeq
where the last term is the two-loop contribution. We have here used the following property of the polylogarithmic functions:
\beq
\frac{d}{dx}\Li_n(x) = \frac{1}{x}\Li_{n-1}(x)
\eeq
Expanding in powers of $\beta\mu$, we find that this agrees with the result of Popov \cite{Popov}. Again we want to write the pressure in terms of the particle density which now becomes
\beq
\rho &=& \left(\frac{mT}{2\pi}\right)^{3/2}\Li_{3/2}(\td{z}) + 4\lambda\left(\frac{m}{2\pi}\right)^3T^2\Li_{1/2}(z)[\Li_{3/2}(z) - \Li_{3/2}(\td{z})]
                                                               \label{eos.5}
\eeq
Inverting this relation, we find to lowest order in $\lambda$
\beq
\mu = T\ln \Li^{-1}_{3/2}(\rho\Lambda_T^3) + 4\lambda\rho
                                                               \label{eos.6}
\eeq
where $\Li_n^{-1}$ is the inverse of $\Li_n$ and $\Lambda_T = \sqrt{2\pi/mT}$ is the thermal wave length. To the same order the self energy is $\Pi = -4\lambda\rho$. Thus, the effective
chemical potential becomes
\beq
\td{\mu} = \mu + \Pi = T\ln \Li^{-1}_{3/2}(\rho\Lambda_T^3)
                                                               \label{eos.7}
\eeq
which equals the usual chemical potential for a free gas. Let us again refer to Fig. 9 where the chemical potentials $\td{\mu}$ and $\mu$ are plotted as function of temperature. The difference $\mu - \td{\mu} = 4\lambda\rho$ found just below the critical temperature, remains constant in the normal phase where the potentials have the same temperature dependence.

As a function of density the pressure becomes
\beq
P = \frac{4\pi}{m}a\rho^2 + \rho T\left[1 - \frac{\Lambda_T^3\rho}{2^{5/2}}\right] + {\cal O}(\rho^3)
                                                               \label{eos.8}
\eeq
The corresponding second virial coefficient reads
\beq
B_2 =-\frac{\Lambda_T^3}{2^{5/2}}\left[1 -8a\sqrt{\frac{mT}{\pi}}\right]
                                                               \label{eos.10}
\eeq
From the relation ${\cal E} = -\del/\del\beta(\beta P)|_{\beta\mu}$ we similarly have for the energy density
\beq
{\cal E} = \frac{4\pi}{m}a\rho^2 + \frac{3}{2}\rho T\left[1 - \frac{\Lambda_T^3\rho}{2^{5/2}}\right] + {\cal O}(\rho^3)
                                                               \label{eos.9}
\eeq
again in agreement with earlier results \cite{LeeYang_3,Huang,Anne}. At $T_C$ these expressions coincide with the limits taken from sub-critical temperatures, found in the last subsection. Summing up, the Eqs. (\ref{eff.11}), (\ref{eff.20}) and (\ref{eos.8}) constitute the equation of state for the Bose gas at all temperatures. 

\section{Discussion and conclusion}
The non-interacting  Bose-Einstein gas in the condensed phase is not a thermodynamically
stable system since its compressibility is infinite \cite{Huang}. Also, the phase transition
from the normal phase is special in that the correlation length diverges below 
the transition temperature. 
This unphysical behavior is directly related to the absence of a real
spontaneous breakdown of the $U(1)$ symmetry of the system with the accompanying Goldstone
bosons which here are the phonon excitations.

With the introduction of a weak repulsion between the particles, the physics of the system
is well-defined in both phases. We have here described the system using modern
field-theoretic methods based on functional integrals. Instead of using the standard
formalism with complex fields as used in the condensed matter literature \cite{FW},
we find it more convenient to use real fields. The divergences in the loop integrals have
been regulated by introducing a physical cutoff. Since these are proportional to powers of
the cutoff, we could instead have used dimensional regularization where they consequently
would be absent. 

We have put special emphasis on
enforcing the consequences of Goldstone's theorem. In the real-field formalism the
important Pines-Hugenholtz theorem then takes a slightly different form. The thermodynamics 
is most directly obtained from
the effective potential whose minimum gives the free energy of the system. At non-zero
temperatures we find that the one-loop effective potential does not give a consistent
description of the symmetry breakdown and that it must be improved by adding in ring
corrections in a self-consistent way. It then becomes natural to introduce an effective
chemical potential $\td{\mu}$ which acts as an order parameter, being positive below 
the critical temperature $T_C$ and giving a non-zero condensate, and passing through zero 
at $T_C$. In this way we find $\td{\mu}$ is a more important variable in characterizing the
phases of the system than the thermodynamic chemical potential $\mu$. In particular, for
temperatures below $T_C$ where the particle number concept looses some of its meaning, the physical significance of $\mu$ is not clear.

Our results are derived to lowest order in the coupling between the particles. As such, most
of the thermodynamic results are already in the literature dating back to the pioneering
calculations of Lee, Yang and collaborators obtained by methods from statistical mechanics.
What is new in addition to the systematic use of the effective chemical potential, is 
basically a more coherent derivation in a framework which has to a large extent been 
developed for corresponding relativistic systems in high energy physics. There is nothing 
preventing us in extending the calculations to higher orders in the interactions. One is 
then forced to consider the theory as an effective field theory which is non-renormalizable
at very high energies, the coupling constant having dimensions of a length. As recently
shown by Braaten and Nieto \cite{BN}, the zero-temperature ground state energy to next order 
in the coupling constant can then be obtained in a most direct way using the renormalization 
group for this effective field theory.

With the recent introduction of magnetic traps, the experimental study of the thermodynamics
of weakly interacting bosons have entered a new phase which will allow a much more detailed
study of this important system. Here we have considered the particles in an open volume
where the interactions dominate, while it is the confining potential which has
the most important r\^ole in the trapped systems \cite{BEC}. The number of particles
in the system now is finite, but one can still use quantum field theory in the grand canonical
ensemble \cite{canonical}. Even with the large effort going on in the field at present, there is
still a long way between experimental results and detailed verifications of the theoretical
properties as derived for the open system here.\\\\
{\large \bf Acknowledgment}\\
We thank Mark Burgess for many useful discussions on ring-corrections of effective potentials at finite temperature.

\appendix
\section{Appendix}
Here we give a brief review of the effective action for composite operators introduced by Cornwall, Jackiw and Tomboulis \cite{CJT}. The method was originally presented in real time formalism, but we will use imaginary time,
following \cite{ACPi}. A compact notation with $\int_x \equiv \int d^dx$ and $J_x\equiv J(x)$ is used. One starts by defining the generating functional
\begin{eqnarray}
Z[J,K] &=& \exp(-W[J,K])\nonumber\\
&=& \int D\phi \exp\left[-\left(S[\phi] + \int_x\phi_xJ_x + \frac{1}{2}\int_{xy}\phi_xK_{xy}\phi_y\right)\right]
                                                               \label{comp.1}
\end{eqnarray}
where the field(s) $\phi$ is coupled both to a local source $J_x$ and
a bilocal source $K_{xy}$. $S[\phi]$ is the classical Euclidean action. 
The background field $\Phi$ and the propagator $G$ are defined by
\begin{eqnarray}
\frac{\delta W[J,K]}{\delta J_x} = \Phi_x
                                                               \label{comp.2}
\end{eqnarray}
\begin{eqnarray}
\frac{\delta W[J,K]}{\delta K_{xy}} = \frac{1}{2}[\Phi_x\Phi_y + G_{xy}]
                                                               \label{comp.3}
\end{eqnarray}
A double Legendre transform then produces the effective action
\begin{eqnarray}
\Gamma[\Phi,G] = W[J,K] - \int_x\Phi_xJ_x - \frac{1}{2}\int_{xy}[\Phi_xK_{xy}\Phi_y + G_{xy}K_{yx}]
                                                               \label{comp.4}
\end{eqnarray}
where $J$ and $K$ are eliminated using (\ref{comp.2}) and (\ref{comp.3}). 
The sources are then related to $\Gamma[\Phi,G]$ via the equations
\begin{eqnarray}
\frac{\delta\Gamma[\Phi,G]}{\delta\Phi_x} = - J_x - \int_yK_{xy}\Phi_y
                                                               \label{comp.5}
\end{eqnarray}
\begin{eqnarray}
\frac{\delta\Gamma[\Phi,G]}{\delta G_{xy}} = - \frac{1}{2}K_{xy}
                                                               \label{comp.6}
\end{eqnarray}
In order to establish a series expansion for the effective action one must
shift the field in the classical action $S[\phi] \rightarrow S[\Phi + \phi]$.
From this the free propagator in background field 
\begin{eqnarray}
D^{-1}_{xy}(\Phi) = \frac{\delta^2 S[\Phi + \phi]}{\delta\phi_x\delta\phi_y}|_{\phi 
= 0}
                                                               \label{comp.7}
\end{eqnarray}
is obtained. A careful rearrangement of terms leads to the series \cite{CJT}
\begin{eqnarray}
\Gamma[\Phi,G] = S[\Phi] + \frac{1}{2}\Tr\ln D_0G^{-1} + \frac{1}{2}\Tr[D^{-1}G - 1] + \Gamma_2[\Phi,G]
                                                               \label{comp.8}
\end{eqnarray}
$\Gamma_2[\Phi,G]$ is the sum of all two-particle-irreducible graphs
obtained from the interaction terms in $S[\Phi + \phi]$ with propagator $G$
and $D_0 = D(\Phi = 0)$
The conventional effective action is now obtained from the equations
\begin{eqnarray}
\Gamma[\Phi] = \Gamma[\Phi,G_0]
                                                               \label{comp.9}
\end{eqnarray}
\begin{eqnarray}
\frac{\delta\Gamma[\Phi,G]}{\delta G}|_{G = G_0} = 0
                                                               \label{comp.10}
\end{eqnarray}
As seen from (\ref{comp.6}) this is equivalent to letting $K = 0$.\\

We now want to compare the effective action (\ref{comp.8}) to the results obtained in the previous sections. $\Phi$ is then the background field
$(v,0)$ and $\phi = (\chi_1,\chi_2)$. We only need $\Gamma$ to 
${\cal O}(\lambda)$. Using $G^{-1} = D^{-1} - \Pi$ we find for the various terms
\begin{eqnarray}
\frac{1}{2}\Tr[D^{-1}G - 1] = \frac{1}{2}\Tr D\Pi + {\cal O}(\lambda^2)
                                                               \label{comp.11}
\end{eqnarray}
\begin{eqnarray}
\frac{1}{2}\Tr\ln G^{-1} = \frac{1}{2}\Tr[\ln D^{-1} - D\Pi + \cdots]
                                                               \label{comp.12}
\end{eqnarray}
Hence, to this order, the effect of the third term in (\ref{comp.8}) is to cancel the ${\cal O}(\lambda)$ contribution, in terms of the propagator $D$ in the one-loop term. To the same order $\Gamma_2$ is simply the sum of two-loop graphs with propagator $G$. Omitting a constant due to $\Tr\ln D_0$ we may to this order write
\begin{eqnarray}
\Gamma[v,\Pi] = S[v] + \frac{1}{2}\Tr\ln G^{-1} - \frac{1}{2}\Tr D\Pi
+ \mbox{two-loop}
                                                               \label{comp.13}
\end{eqnarray}
If we now make the ansatz for $G = [D^{-1} - \Pi]^{-1}$, as we did assuming the Goldstone theorem, this is seen to equal our result for the ring-corrected effective potential.

\end{document}